\documentclass[a4paper,fleqn]{cas-sc}

\usepackage[authoryear]{natbib}
\usepackage{placeins}
\usepackage{graphicx} 
\usepackage{subcaption} 
\usepackage[authoryear]{natbib}
\usepackage{tabularx}
\usepackage{changepage}
\usepackage{subcaption}
\usepackage{multicol}
\usepackage{rotating}
\usepackage{changepage}

\def\tsc#1{\csdef{#1}{\textsc{\lowercase{#1}}\xspace}}
\tsc{WGM}
\tsc{QE}
\tsc{EP}
\tsc{PMS}
\tsc{BEC}
\tsc{DE}

\usepackage{framed} 

\usepackage{multicol} 

\usepackage{nomencl} 
\DeclareUnicodeCharacter{03B2}{\ensuremath{\beta}}
\makenomenclature

\renewcommand*\nompreamble{\begin{multicols}{2}}

\renewcommand*\nompostamble{\end{multicols}}

\begin{document}

\let\WriteBookmarks\relax
\def\floatpagepagefraction{1}
\def\textpagefraction{.001}

\shorttitle{}

\shortauthors{Rebello, C. M. et~al.}

\title{Optimizing $CO_2$ Capture in Pressure Swing Adsorption Units: A Deep Neural Network Approach with Optimality Evaluation and Operating Maps for Decision-Making}



%
\author[1]{Carine Menezes Rebello}[
                        orcid=0000-0002-0796-8116]



\ead{carine.m.rebello@ntnu.no}


\author[1]{ Idelfonso B. R. Nogueira}[
                        orcid=0000-0002-0963-6449]

\affiliation[1]{organization={Department of Chemical Engineering},
    addressline={Norwegian University of Science and Technology}, 
    city={Trondheim},
    postcode={793101}, 
    country={Norway}}
\cormark[1]
\cortext[cor1]{Corresponding author}



\begin{abstract}
This study presents a methodology for surrogate optimization of cyclic adsorption processes, focusing on enhancing Pressure Swing Adsorption units for carbon dioxide ($CO_2$) capture. We developed and implemented a multiple-input, single-output (MISO) framework comprising two deep neural network (DNN) models, predicting key process performance indicators. These models were then integrated into an optimization framework, leveraging particle swarm optimization (PSO) and statistical analysis to generate a comprehensive Pareto front representation. This approach delineated feasible operational regions (FORs) and highlighted the spectrum of optimal decision-making scenarios. A key aspect of our methodology was the evaluation of optimization effectiveness. This was accomplished by testing decision variables derived from the Pareto front against a phenomenological model, affirming the surrogate model's reliability. Subsequently, the study delved into analyzing the feasible operational domains of these decision variables. A detailed correlation map was constructed to elucidate the interplay between these variables, thereby uncovering the most impactful factors influencing process behavior. The study offers a practical, insightful operational map that aids operators in pinpointing the optimal process location and prioritizing specific operational goals.

\end{abstract}


\begin{highlights}
\item Enhancing PSA efficiency with AI-based multi-objective optimization.
\item The validation of optimization through artificial neural networks is important to assess the reliability of the optimized process.
\item Understanding the viable operating region (FOR) helps identify the practical limits within which the system operates effectively and efficiently.
\end{highlights}


\maketitle

\begin{table}[h!]   

\begin{framed}
\footnotesize

\nomenclature{$K$}{Compact subset of $\mathbb{X}$}
\nomenclature{$H$}{Generic prediction function or Enthalpy [J/mol]}
\nomenclature{$A$}{Area of column [m$^2$]}
\nomenclature{$\omega$}{Prediction uncertainty}
\nomenclature{$ln$}{Natural logarithm}
\nomenclature{$\mathbf{\theta}$}{Decision variables in Fisher test}
\nomenclature{$\lambda$}{Vector of Lagrange multipliers in Fisher test}
\nomenclature{$n_{\mathbf{\theta}}$}{Number of decision variables}
\nomenclature{$\alpha$}{Confidence level of the Fisher–Snedecor test}
\nomenclature{$\varepsilon$}{Predetermined tolerance}
\nomenclature{$k$}{Order of the function or order parameters}

\nomenclature{$L$}{Fisher test function or length of column [m]}
\nomenclature{$P $}{ Pressure [Pa]}
\nomenclature{$P_0 $}{ Adsorption pressure [Pa]}
\nomenclature{$P^* $}{ Dimensionless pressure}
\nomenclature{$T $}{ Temperature [K]}
\nomenclature{$T_0 $}{ Feed temperature [K]}

\nomenclature{$x_i $}{ Dimensionless molar loading of component $i$ in the solid phase}
\nomenclature{$q_i $}{ Molar loading of component $i$ in the solid phase [mol/kg]}
\nomenclature{$q_{s0}$}{ Molar loading scaling factor [mol/kg]}
\nomenclature{$\nu_0 $}{ Velocity scaling factor [m/s]}
\nomenclature{$\nu_z $}{ Superficial gas velocity [m/s]}
\nomenclature{$\nu^*_z $}{ Dimensionless superficial gas velocity}
\nomenclature{$Z $}{ Dimensionless length coordinates}
\nomenclature{$z $}{ Bed length coordinate [m]}
\nomenclature{$x_i^* $}{Dimensionless equilibrium molar loading of component
$i$ in the solid phase}
\nomenclature{$q_i*$}{ Equilibrium molar loading of component $i$ in the solid
phase [mol/kg]}
\nomenclature{$R $}{ Universal gas constant [J/(mol.K)]}
\nomenclature{$D$}{Inner diameter of column [m]}
\nomenclature{$D_m$}{Molecular diffusivity of $CO_2-N_2$ mixture [m$^2$/s]}
\nomenclature{$D_L$}{Axial dispersion coefficient [m$^2$/s]}
\nomenclature{$r_p$}{ Radius of adsorbent pellet [m]}

\nomenclature{$\tau$}{Dimensionless time}

\nomenclature{$C_{p,a}$}{Specific heat capacity of the adsorbed phase [J/(mol.K)]}
\nomenclature{$C_{p,g}$}{Specific heat capacity of the gas [J/(mol.K)]}
\nomenclature{$C_{p,s}$}{Specific heat capacity of the adsorbent [J/(kg.K)]}
\nomenclature{$C_{p,w}$}{Specific heat capacity of the adsorbent [J/(kg.K)]}
\nomenclature{$k_i$}{Mass transfer coefficient of component i [s$^{-1}$] or Total number of predictions by SciML}
\nomenclature{$h_{in}$}{Inner heat transfer coefficient [W/(m$^2$.K)]}
\nomenclature{$h_{out}$}{Outer heat transfer coefficient [W/(m$^2$.K)]}
\nomenclature{$K_w$}{ Thermal conductivity of the wall [W/(m.K)]}
\nomenclature{$K_z$}{ Effective gas thermal conductivity [W/(m.K)]}
\nomenclature{$L_{wf}$}{ Length of column affected by the waterfront [m]}
\nomenclature{$MW_i$}{ Molecular weight of component $i$ [kg/mol]}
\nomenclature{$N$}{ Number of columns}
\nomenclature{$r_i$}{Column inner radius [m]}
\nomenclature{$r_o $}{Column outer radius [m]}
\nomenclature{$t $}{ Time}
\nomenclature{$T_a $}{ Ambient temperature [K]}
\nomenclature{$T_0 $}{ Feed temperature [K]}
\nomenclature{$T_w $}{ Column wall temperature [K]}
\nomenclature{$T^*$}{ Dimensionless temperature}
\nomenclature{$T^*_{w} $}{ Dimensionless wall temperature}
\nomenclature{$U$}{ Internal energy [J/mol]}
\nomenclature{$w_l $}{ Water mole fraction in the feed gas to PSA unit}
\nomenclature{$y_i $}{ Mole fraction of component $i$ in the gas phase}

\nomenclature{$L$}{Lagrangian function, [-]}
\nomenclature{$\lambda$}{Vector of Lagrange multipliers, [-]}
\nomenclature{$\boldsymbol{\theta}_j^*$}{Optimal vector of operating conditions, [-]}
\nomenclature{$\boldsymbol{\theta}_k$}{Vector of particles, [-]}
\nomenclature{$n_i$}{Number of objective functions, [-]}
\nomenclature{$n_k$}{Number of particles, [-]}
\nomenclature{$n_j$}{Point number on the Pareto front, [-]}
\nomenclature{$n_{\theta}$}{Number of decision variables, [-]}
\nomenclature{$F$}{Test Fisher, [-]}

\printnomenclature

\end{framed}

\end{table}

\section{Introduction}
Cyclic adsorption processes are recognized for efficiently separating complex mixtures while presenting energy efficiency and environmental sustainability \citep{NOGUEIRA2020107821}. Among these processes, the pressure swing adsorption (PSA) unit stands out for its diverse range of applications in gas-phase separations. These applications span various industries, including the petrochemical sector \citep{DENG201719984,DOBLADEZ2020106717,Pruksathorn2009, Jeong}, $CO_2$ capture \citep{Haghpanah2013, Krishnamurthy2014, Leperi,SIQUEIRA20172182,Arvind, PAI2020116651}, and the fine chemicals field \citep{Dong}. Nevertheless, the PSA model exhibits complex dynamics characterized by cyclic behavior instead of steady-state operation. This occurs due to the alternation between the different stages that define the system, which continually vary throughout the cycle. Furthermore, PSA systems often incorporate multiple adsorption beds operating in series or parallel, each with individual characteristics. The quest for energy optimization in pressure swing adsorption (PSA) modeling and the necessity to account for transient effects and multiphase transport presents a challenge. This demands the development of sophisticated approaches that accurately represent the complex behavior of PSA processes and enhance their efficiency. Crucially, these approaches must be designed to optimize process performance while ensuring minimal computational effort, striking a balance between detailed representation and practical feasibility. 

Building on the challenges highlighted in the pursuit of energy optimization in PSA modeling, the operation of Pressure Swing Adsorption itself is inherently complex due to the continuous alternation of steps in the process. This complexity is further compounded when trying to achieve efficient operation, as it requires the precise definition and coordination of various operational variables, including pressures, temperatures, gas flow rates, and cycle times. Moreover, PSA processes often involve conflicting performance metrics \citep{HAO2021130248, ZHU20181061}, where optimizing one parameter can adversely affect another \citep{Witte,Lorenz}. This interplay of variables and the trade-offs involved underscore the need for advanced optimization strategies that can navigate these complexities while maintaining process efficiency and low computational burden.  For instance, pursuing higher purity and recovery may lead to trade-offs in the process, making optimizing the PSA process a significant challenge, requiring careful balance among various operational factors to achieve the desired performance and overall efficiency \citep{ZHANG20215403}.

In this scenario, \cite{Barg} focus on the simulation and optimization of an industrial PSA unit for hydrogen purification. The research involves developing a comprehensive mathematical model that accurately represents the dynamics of a commercial PSA unit. This model is then utilized to optimize key operational parameters, such as cycle time and the steps involved in the pressure swing process. \cite{nogueira2020big}  addressed the complexity of optimizing pressure swing adsorption processes. Their work presents a method to optimize a PSA unit for syngas purification using porous aminofunctionalized titanium terephthalate MIL-125. The integrates big data analytics and metaheuristic techniques, aiming to effectively address and manage the uncertainties commonly encountered in the optimization process. This approach highlights the computational challenges of using phenomenological models in PSA process optimization, particularly due to the complexity of the partial differential equation systems.

The difficulties in conducting optimizations based on first-principle models are evident in various fields where rigorous modeling results in complex numerical problems. In such scenarios, the challenge lies in balancing the accuracy of detailed models with the feasibility of computational resources and time constraints.  For PSA processes, \cite{SMITH19912967} addressed this issue by proposing an innovative approach to the cost-optimal design of PSA systems. It introduces a model that integrates design elements, such as the number of beds and operational sequences, to minimize the annualized cost of PSA separation systems. The model simplifies the complex dynamics of PSA processes into algebraic equations, balancing capital and operating costs \citep{SMITH19912967}. On the other hand, \cite{jiang2005design} introduces a Newton-based approach for determining cyclic steady states in PSA systems, integrating design constraints for more accurate optimization. The study also employs advanced sequential quadratic programming (SQP)-based algorithms for designing optimal PSA processes, incorporating detailed adsorption models. A key innovation is the parallelization of sensitivity calculations, significantly improving computational efficiency.  \cite{haghpanah2013multiobjective} highlight these issues focusing on optimizing a four-step adsorption process for $CO_2$ capture using a novel finite volume-based simulator. This research addresses the numerical challenges in adsorption process simulations, comparing various high-resolution schemes for accuracy and efficiency. The study's core contribution lies in its rigorous optimization approach, which efficiently identifies optimal operating conditions for $CO_2$ capture, balancing process demands with energy consumption. This work marks a significant step in advancing numerical methods for optimizing adsorption processes. 

\cite{agarwal2009simulation}, address the computational challenges in simulating and optimizing pressure swing adsorption systems. The authors highlight that optimizing these systems, due to the coupled stiff PDE system and steep fronts moving with time, poses significant computational difficulties for existing optimization techniques. To overcome this, the study develops a reduced-order model (ROM) based on proper orthogonal decomposition (POD). This ROM serves as a low-dimensional approximation to the dynamic PDE-based model, offering a more cost-efficient and manageable approach for the simulation and optimization of PSA systems. 

In this scenario, more recently, artificial intelligence (AI), specifically artificial neural networks (ANN), has been proposed as a surrogate model to reduce the computational effort. This approach facilitates the extraction and storage of vast information from databases while requiring minimal computational resources for simulation. Consequently, ANN models have emerged as a bridge connecting rigorous phenomenology with the low computational effort required for real-time applications \citep{ALKEBSI2021120078,doi:10.1021/acs.iecr.9b02383,NOGUEIRA2017123,SANTANNA2017377,Subraveti,TONG2021100075}. 

For instance, 
 \cite{OLIVEIRA2020115801} introduced a framework for deep neural networks (DNN) models in PSA units, showcasing the advantages of DNNs in capturing intricate system behavior or representing system performance parameters. Addressing the computational effort related to the optimization procedure of these processes and the issue of properly selecting the optimal sorbent,  \cite{nogueira2022novel} presents a strategy for simultaneous material screening and process optimization. This strategy leverages the capabilities of deep neural networks to extract knowledge from databases and employs a nested optimization loop that simultaneously considers process and material perspectives. The single objective optimization problem results are analyzed using a Fisher–Snedecor test, which assesses the uncertainties of optimal points and helps build feasible operating regions for the processes.  \cite{subraveti2022can} introduces an innovative approach using physics-based artificial neural networks (PANACHE) to simulate and optimize chromatographic processes. This method addresses the computational challenges of solving chromatography's hyperbolic PDEs and nonlinear adsorption isotherms. The approach significantly improves the accuracy of process simulations by learning the underlying PDEs through a physics-constrained loss function. The study demonstrates the effectiveness of this method with binary solute mixtures and various isotherm systems, achieving up to 250 times faster computational speeds.  \cite{pai2020generalized} addresses the complexities in simulating and optimizing adsorption-PSA. This study introduces an ANN generalized and adsorbent-agnostic framework, significantly accelerating the simulation and optimization process. The ANN framework facilitates rapid adsorbent screening and process optimization, effectively reducing the computational effort typically associated with these procedures.  \cite{MARTINS2021119333} proposed DNN identification and integration into an economic predictive control scheme, serving as a predictive model within the controller.

 In this context, a series of recent studies have delved into applying Deep Neural Network strategies for optimizing  PSA units. Collectively, these works signify the advent of DNN-based approaches in PSA optimization, showcasing the integration of advanced machine learning techniques in refining the efficiency and effectiveness of PSA processes. This trend highlights a  shift towards leveraging data-driven methodologies in the field of PSA optimization \citep{Leperi,MA20195324,SANTANNA2017377,Subraveti,Xiao2020,YU202111740}.

These surrogate models can be useful tools not only for optimization but also for tasks such as online soft-sensing and predictive modeling for model predictive controllers.  In \cite{pr10020409}, the authors address issues related to PSA complex dynamics. They propose the development of data-driven sensors for the PSA inference problem. The study particularly focuses on addressing the uncertainties inherent in these sensors, proposing a novel approach using soft sensors and deep learning techniques. In \cite{COSTA2024107364}, the authors extend their DNN-based approach, developing an adaptive digital twin for pressure swing adsorption systems based on neural network models. This research likely explores integrating a novel feedback-tracking system with online learning and uncertainty assessment to improve the performance of PSA systems. In this context, a digital twin suggests a sophisticated simulation model that mirrors the real-world PSA system, allowing for real-time monitoring, prediction, and optimization. 

While the aforementioned studies have significantly advanced the literature, a common limitation is the lack of systematic evaluation of optimization optimality. While the advantages of ANNs in these contexts are clear, assessing their accuracy in achieving the optimization tasks is crucial. This is particularly important considering that surrogate-based optimization, like that involving ANNs, can sometimes introduce artificial minima into the optimization landscape. Such occurrences can mislead the optimizer, potentially yielding unreliable results. Therefore, failing to evaluate these models rigorously could lead to representations that do not accurately reflect the system's behavior in practical scenarios, underscoring the need for thorough validation of ANN-based optimization strategies. 

Tackling the optimization task while simultaneously identifying a feasible operation region (FOR) remains a complex yet crucial challenge in surrogate-based optimization for PSA processes. This innovative concept, initially introduced by \cite{NOGUEIRA2020107821} for true moving bed reactors, presents a valuable strategy. The FOR approach generates a 'cloud' of optimal solutions, from which a subset can be selected based on specific criteria. These selected points can undergo experimental validation or rigorous model testing, comprehensively evaluating the effectiveness of the optimization. Implementing this methodology offers several benefits, particularly in enhancing understanding the interactions between various operational variables in intricate, multi-stage processes like PSA. Moreover, the FOR concept proves highly beneficial in real-time applications, where the accuracy and reliability of these models are important. This approach not only aids in ensuring the precision of surrogate models but also contributes significantly to the field of PSA process optimization. 

In this study, we introduce a novel framework for the multi-objective optimization of pressure swing adsorption units, utilizing deep neural network models and incorporating an assessment of optimality through feasible operation regions. This framework integrates an operations map with the Pareto front, from which samples are selected and rigorously tested against a detailed model to confirm the optimization's effectiveness. This approach significantly reduces the computational effort compared to traditional PDE-based optimization methods while also ensuring the reliability of the surrogate-based optimization results. The key contributions of this work are twofold: firstly, it presents an innovative method for optimizing PSA processes using DNNs, aligned with a Pareto-optimal FOR front; secondly, it provides a comprehensive evaluation of the optimality of DNN-based optimization strategies. Through this study, we aim to enhance the efficiency and accuracy of PSA unit optimization, particularly in the context of carbon capture applications.

\section{Methodology}

The first step of the methodology proposed in this study is obtaining the dataset. These datasets are then fed to the surrogate model identification framework. In sequence, a multi-objective optimization proposed is solved using the surrogate models to identify the Pareto Front. Then, the points in the Pareto front are tested to evaluate the optimality of the surrogate-based optimization.

\begin{figure}[t]
\centering
\includegraphics[width=16cm]{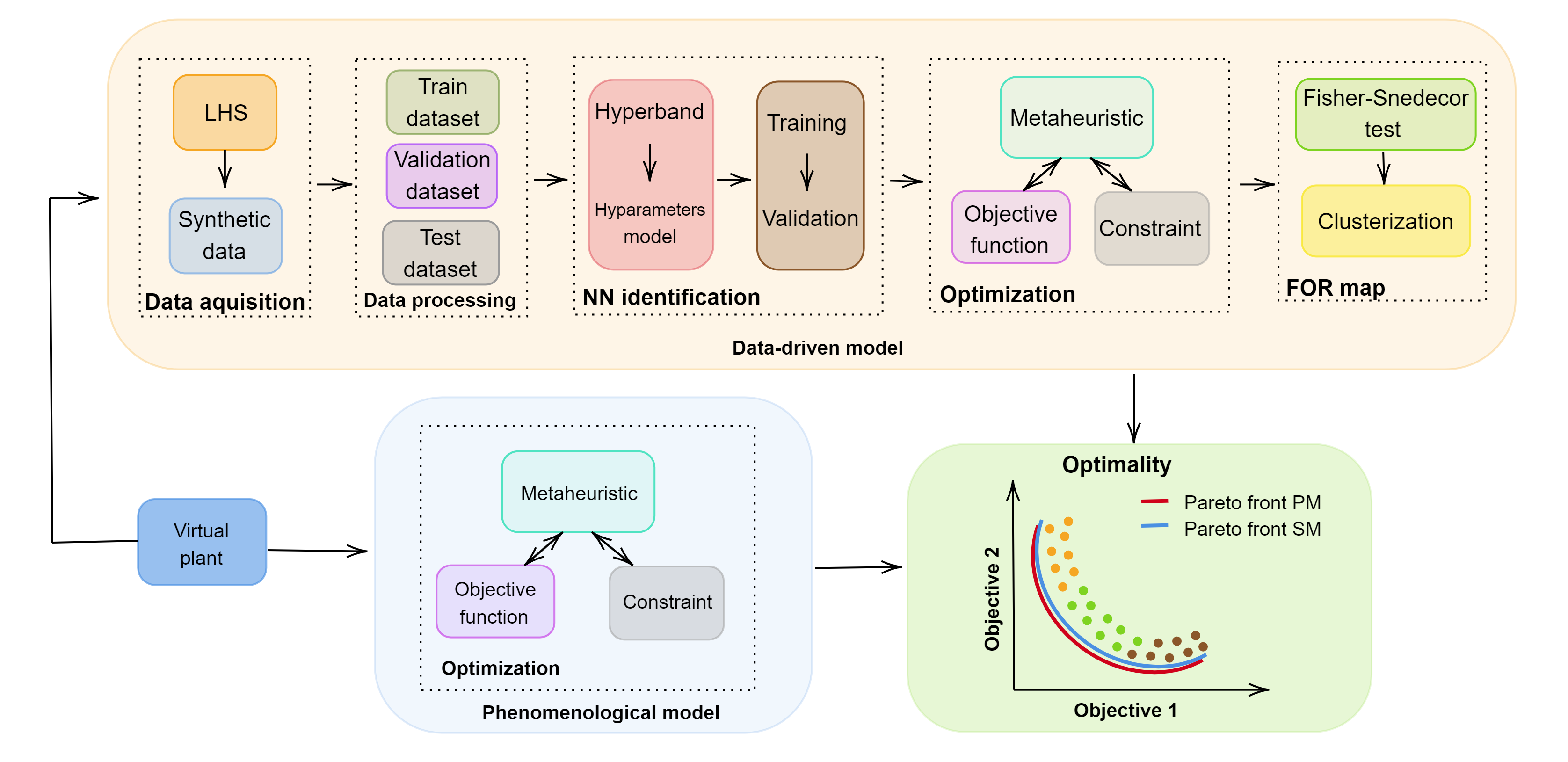}
\caption{Diagram illustrating the methodological steps developed in this work.\label{medologia}}
\end{figure}

The Fisher-Snedecor test establishes a feasible operational region near the Pareto front. This operational region is subsequently divided into three distinct subgroups, forming the core components for constructing the operational map, as depicted in Figure \ref{medologia}. Finally, a detailed process analysis is carried out for the three different regions. Then, the operation map is defined. Further elaboration of these details is provided below.

\subsection{Pressure Swing Adsorption to $CO_2$ Capture}

There has been a substantial increase in atmospheric $CO_2$ levels, with an emerging scientific consensus indicating that this rise is a primary contributor to recent global warming. This upward trend in global temperatures has been a focal point in the contemporary climate change discussion. In response, continuous effort has been to develop and implement renewable energy technologies to minimize carbon emissions. Despite these advancements, fossil fuels will remain relevant to the energy matrix for some time. Reducing $CO_2$ emissions has become a key focus in this context, with special attention to techniques such as carbon capture and storage (CCS). 
For this purpose, a technological alternative that has attracted interest in recent years is separation by cyclic adsorption. Among these techniques, pressure swing adsorption is renowned for its operation that balances cost-efficiency with effectiveness. Particularly, it demonstrates proficiency in facilitating challenging gas phase separations. Operated cyclically, these systems achieve a pseudo-continuous performance where each bed recurrently operates through a series of different steps. The effectiveness of this technology, in terms of purity and recovery capacity, is crucial for optimizing costs and ensuring efficient capture of $CO_2$ emissions.

In this work, PSA system employing the Skarstrom cycle \citep{Leperi,Ebner2009} was adopted for the separation of $N_2$ and $CO_2$, utilizing zeolite 13X as the adsorbent material. Zeolite 13X is recognized as one of the high-performance adsorbents in the field of gas separation and purification. Characterized by its unique microporous structure and high adsorption capacity, this zeolite stands out in the selective capture of specific molecules, such as $N_2$ and $CO_2$ \citep{Leperi}. This process configuration is widely used in PSA due to its simplified operation and long history of industrial use.  This cycle consists of the following stages: (1) pressurization (Pres), (2) adsorption (Ads), (3) heavy reflux (HR), (4) counter-current depressurization (Depres), and (5) light reflux (LR).

In the initial pressurization stage, a gas mixture comprising $CO_2$ and $N_2$ elevates the pressure of the column. In the subsequent adsorption stage, $CO_2$ gets adsorbed in the bed, while $N_2$ is produced. As the bed approaches saturation, an additional high-purity $CO_2$ stream is introduced into the column to purge residual $N_2$ within the bed (HR stage). Following this in the counter-current depressurization stage, the column undergoes depressurization to desorb the  $CO_2$ from the bed, transitioning from a high-pressure state to a low-pressure state. In the light reflux stage, high-purity $N_2$ is injected into the bed to regenerate the column by removing part of the remaining $CO_2$.

In a PSA process, a wide range of operating conditions must be specified throughout the cycles to ensure high purity and recovery of the separated gases. The set of variables may include pressure, feed temperature, step durations, and flow rates, making the optimization of this unit challenging. This set of decision variables must be adjusted to meet operation targets while satisfying purity and recovery constraints. 

On the other hand, several performance metrics must be evaluated. The purity ensures that the final product meets the required quality standards.
The recovery metric is important because it determines how much the desired component is recovered from the process, minimizing losses and optimizing efficiency. Due to complex dynamics, obtaining high purity and recovery might not be a simple task.

\begin{figure}[t]
\includegraphics[width=14 cm]{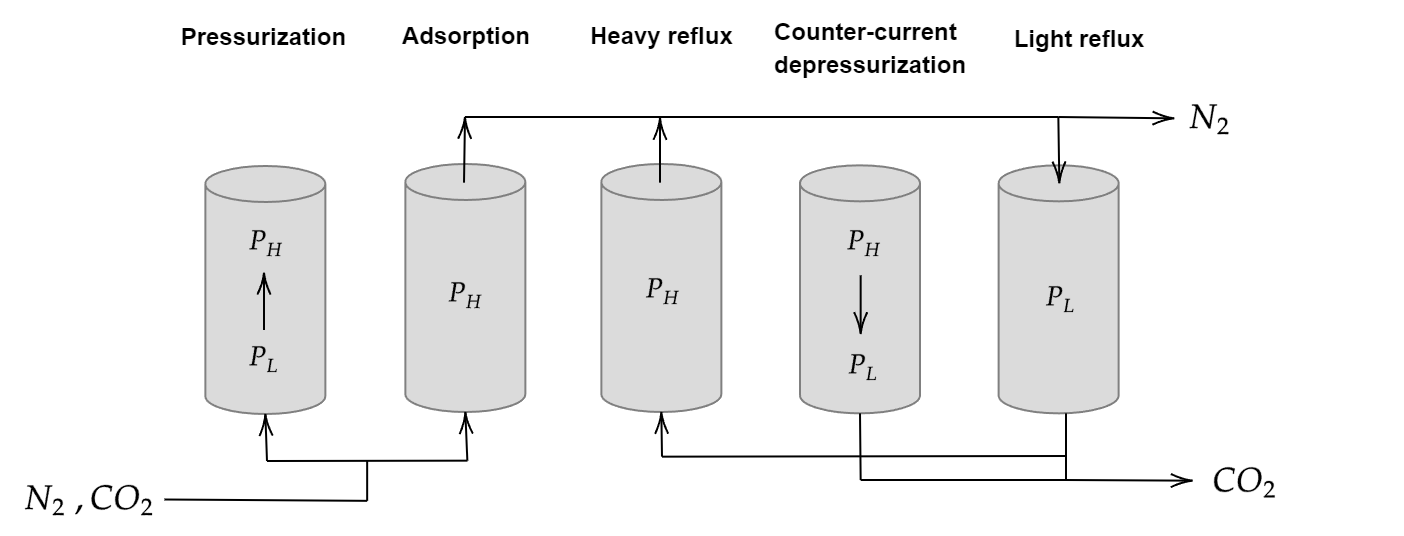}
\caption{Illustrative diagram of the pressure swing adsorption steps used in this study}
\label{PSA_esquema}
\end{figure}

\subsection{Phenomenological Model} \label{model}

In this work, the above PSA was modeled by a system of partial differential equations (PDEs) and nonlinear algebraic equations to represent the system's mass, energy, and momentum balances \citep{Leperi}. Some assumptions were considered, as shown below:

\begin{itemize}
    \item[-] Ideal gas law is used to describe the gas phase;
    \item[-] No concentration, pressure, or temperature gradient in the radial or azimuth directions;
    \item[-] Linear driving force is used to describe the gas diffusion into the adsorbent;
    \item[-] Adsorbent properties and void fraction are constant throughout the column;
    \item[-]  Viscosity of the gas is independent of pressure;
    \item[-] There is thermal equilibrium between the adsorbent and the gas phase;
    \item[-] The Ergun equation is used to describe the pressure drop across the bed;
    \item[-] The column operates adiabatically; wall energy balance is not needed;
    \item[-] An axially dispersed plug flow model represents bulk fluid flow.
\end{itemize}

All of the equations employed in the model were converted into dimensionless forms. When working with dimensionless quantities, simulations' numerical stability and accuracy are improved, mitigating problems associated with extremely large or small numbers. The dimensionless variables are provided below:

\begin{align}
&P^*=\frac{P}{P_0},\; T^*=\frac{T}{T_0}, x_{i\ }=\frac{q_i}{q_{s0}}, { \nu^*}_Z=\frac{\nu_Z}{v_0}, 
\tau=\frac{t\nu_0}{L},
x_i^\ast=\frac{q_i^\ast}{q_{s0}}, Z\ =\frac{z}{L}
\label{adimensionalizacao}
\end{align}

\noindent where $P$ is the pressure of the system, $P^*$ is dimensionless pressure, $P_{0}$ is adsorption pressure, $T^*$ is dimensionless temperature, $T$ is temperature, $T_{0}$ is feed temperature, $x_i$ is dimensionless molar loading of component $i$ in the solid phase, $q_i$ is molar loading of component $i$ in the solid phase, $q_{s0}$ is molar loading scaling factor, $v_Z$ is superficial gas velocity, $ \nu^*_Z$ dimensionless superficial gas velocity, $v_0$ is velocity scaling factor, $\tau$ is dimensionless time, $t$ is time, $L$ is the length of the column, $x_i^\ast$ is dimensionless equilibrium molar loading of component $i$ in the solid phase, $q_i^\ast$ is the equilibrium molar loading of component $i$ in the solid phase, $q_{s0}$ is the molar loading scaling factor, $Z$ is the dimensionless length coordinate, $z$ is bed length coordinates. 

The mass balance (Equation \ref{mass_balance}) implemented in this model was computed about to the molar fraction of $CO_2$ (represented by $y_i$) in the gas phase. This mass balance is calculated by:

\begin{align}
\frac{\partial y_i}{\partial\tau}=\frac{1}{Pe}\left(\frac{\partial^2y_i}{\partial Z^2}+\frac{1}{P^*}\frac{\partial P^*}{\partial Z}\frac{\partial y_i}{\partial Z}-\frac{1}{T^*}\frac{\partial T^*}{\partial Z}\frac{\partial y_i}{\partial Z}\right)  -{\bar{v}}_Z\frac{\partial y_i}{\partial Z}  +\frac{\Psi T^*}{P^*}\left(\left(y_i-1\right)\frac{\partial x_i}{\partial\tau}+y_i\frac{\partial x_{i+1}}{\partial\tau}\right)
\label{mass_balance}    
\end{align}

\noindent where $\Psi\ =\left[\frac{(1-\varepsilon)}{\varepsilon}\right]\ \left(\frac{RT_0q_{s0}}{P_0}\right)$, and the Peclet number is calculated by:
\begin{equation}
P_e=\frac{v_0L}{D_L}
\label{Peclet}
\end{equation}

The axial dispersion coefficient, $D_L$, is calculated using the following: 

\begin{equation}
D_L=0.7D_m+r_p\cdot v_0
    \label{eq_DL}
\end{equation}

\noindent where $D_m$ is the molecular diffusivity of $CO_2$ 
 - $N_2$ mixture and $r_p$ is the radius of the adsorbent pellet.
The total system pressure was calculated using the bed total mass balance equation:
\begin{align}
\frac{\partial P^*}{\partial\tau}=&\left(-P^*\frac{\partial\left({\bar{v}}_Z\right)}{\partial Z}-{\bar{v}}_Z\frac{\partial\left(P^*\right)}{\partial Z}+{\bar{v}}_Z\frac{P^*}{T^*}\frac{\partial\left(T^*\right)}{\partial Z}\right) \Psi T^*\sum_{i}\frac{\partial\bar{x_i}}{\partial\tau}+\frac{P^*}{T^*}\frac{\partial T^*}{\partial\tau}
    \label{balanco_massa}
\end{align}

The mass transfer between the gaseous and solid phases is calculated by the linear driving force (LDF) given by:
\begin{equation}
\frac{\partial x_i}{\partial\tau}=\frac{k_iL}{v_0}(x_i^\ast-x_i)
    \label{balanco_solido}
\end{equation}
\noindent where $k_i$ is the mass transfer coefficient of component $i$.

The pressure drop throughout the column is calculated using the Ergun equation:

\begin{align}
-\frac{\partial P^*}{\partial Z}=&\frac{150\mu\left(1-\varepsilon\right)^2L}{4r_p^2\varepsilon^2P_0}{\bar{v}}_Zv_0 +\frac{1.75\left(1-\varepsilon\right)L}{2r_p\varepsilon P_0}\left(\sum_{i} y_iMW_i\rho_g\right){{\bar{v}}_Z}^2v_0^2
    \label{ergun}
\end{align}

\noindent where $\mu$ is gas viscosity, $\varepsilon$ is bed void fraction, $r_p$ is radius of adsorbent pellet and $MW_i$ is molecular weight of component $i$. 

The density of the gas phase ($\rho_g$) is given by:
\begin{equation}
\rho_g=\frac{P^*P_H}{RT_0T^*}
    \label{massa_esp}
\end{equation}

\noindent where $R$ is the universal gas constant and $P_H$ is the high pressure.

The energy balance equation for the gas phase is defined as follows:
\begin{equation}
\frac{\partial T^*}{\partial\tau}=\ \pi_1\frac{\partial^2T^*}{\partial\tau^2}-\ \pi_2{\bar{v}}_Z\frac{\partial T^*}{\partial\tau}+\sum_{i}{(\pi_{3i}+\ \pi_4T^*)\frac{\partial x_i}{\partial\tau}}
    \label{balanco energia}
\end{equation}
\noindent where
\begin{equation}
\pi_1=\frac{K_Z}{\left(\rho_gC_{p,g}\varepsilon+(1-\varepsilon)\ \left(C_{p,s}\rho_s+C_{p,a}q_{s0}\right)\right)v_0L}
    \label{pi1}
\end{equation}

\begin{equation}
\pi_2=\frac{\rho_gC_{p,g}\varepsilon}{\left(\rho_gC_{p,g}\varepsilon+(1-\varepsilon)\ \left(C_{p,s}\rho_s+C_{p,a}q_{s0}\right)\right)}
    \label{p2}
\end{equation}

\begin{equation}
\pi_{3i}=\frac{(1-\varepsilon )(-\Delta H_i)q_{s0}}{T_0 \left ( \rho_g C_{p,g}  \varepsilon + (1-\varepsilon)(C_{p,s} \rho_s + C_{p,a} q_{s,0}) \right )}
    \label{p3}
\end{equation}

\begin{equation}
\pi_4=\frac{(1-\varepsilon)(C_{p,g}-C_{p,a})q_{s0}}{\left(\rho_gC_{p,g}\varepsilon+(1-\varepsilon)\ \left(C_{p,s}\rho_s+C_{p,a}q_{s0}\right)\right)}
    \label{p4}
\end{equation}

\begin{equation}
\Delta H_i =\Delta U_i-RT^*T_0
    \label{Hi}
\end{equation}

\noindent where $C_{p,s}$ is specific heat capacity of the adsorbent, $C_{p,g}$ is specific heat capacity of the gas, $C_{p,a}$ is specific heat capacity of the adsorbed phase, $K_Z$ is effective gas thermal conductivity, $\rho_s$ is density of adsorbent, $\Delta H_i$ is enthalpy change, $\Delta U_i$ is internal energy change, $\pi$ is dimensionless groups in the energy balance equations.

This work considers the zeolite 13X to have two temperature-dependent Langmuir sites. We employed isothermal parameters to depict the adsorption process, as summarized in Table \ref{Isoterma}. The associated isotherm equations, derived from \cite{Haghpanah}, are as follows:

\begin{equation}
q_i^\ast=\frac{q_{sat,i}^bB_iy_iP}{1+\sum_{i}{B_iy_iP}}+\frac{q_{sat,i}^dD_iy_iP}{1+\sum_{i}{D_iy_iP}}
    \label{solido}
\end{equation}
 \noindent in which
\begin{align}
B_i=b_i exp^{\frac{-\Delta U_{b,i}}{RT}},
    \label{parametro_B} \\
D_i=d_i exp^{\frac{-\Delta U_{d,i}}{RT}}.
    \label{Parametro_D}
\end{align}

Table \ref{Tabela_parametros} presents some of the parameters and specifications required to solve the phenomenological model for PSA.

\begin{table}
\renewcommand{\arraystretch}{1.5} 
\caption{Isotherm parameters for $CO_2$ and $N_2$ for zeolite 13X on two sites named b and d.}\label{Isoterma}
\begin{tabular}{lccll}
\cline{1-3}
Parameters                                                               & $CO_2$                                & $N_2$                                & \multicolumn{1}{c}{} & \multicolumn{1}{c}{} \\ \cline{1-3}
Saturation loading for site b, $q^b_{sat,i}(mol/kg)$   & 6.21                               & 11.9                              & \multicolumn{1}{c}{} & \multicolumn{1}{c}{} \\
Saturation loading for site d, $q^d_{sat,i}(mol/kg)$   & 7.15                               & 0.00                              & \multicolumn{1}{c}{} & \multicolumn{1}{c}{} \\
Pre-exponential factor for site b, $b_i (kPa^{-1})$        & $4.47\times10^{-8}$ & $1.40\times10^{-7}$ &                      &                      \\
Pre-exponential factor for site d, $d_i (kPa^{-1})$       &$ 4.71\times10^{-10}$ & 0.00                              &                      &                      \\
Heat of adsorption for site b, $\Delta U_{b,i} (kJ/mol)$ & -37.8                              & -19.4                             &                      &                      \\
Heat of adsorption for site d, $\Delta U_{d,i} (kJ/mol)$ & -37.8                              & 0.00                              &                      &                      \\ \cline{1-3}

\end{tabular}
\end{table}

Table \ref{Contorno} presents the model's initial conditions. As we initiate the simulation, the bed and column are assumed to have reached equilibrium with the gas phase at atmospheric temperature. During the pressurization phase, the same pressure as in the purge phase is maintained. Following the initialization phase, the initial conditions (refer to Equation \ref{Cond_inicial}) for each subsequent phase are presumed to align with the bed profile after the prior phase. This ensures a transition and coherence of the simulation. 

\begin{align}
&P^* = P^*_L,\;    y_i = y_{feed},\;    T^* = T_a,\;   x_i = x_i^*\mid_{y_{feed}}\;   
\label{Cond_inicial}
\end{align}

\begin{table}[]
\caption{PSA model parameters}
\renewcommand{\arraystretch}{1.5}
\begin{tabular}{@{}p{8cm}@{\hspace{3cm}}c@{}}
\hline
Parameters                                                 & Value                 \\ \hline
Length of the column, $L$ $(m)$                                & 1                     \\
Inlet gas $CO_2$ mole fraction, $y_0$                      & 0.15                  \\
Viscosity of gas, $\mu$ $(Pa.s)$                             & $1.72 \times 10^{-5}$ \\
Void fraction, $\varepsilon$                               & 0.37                  \\
Molecular diffusivity, $D_m$ $(m^2/s)$      & $1.3 \times 10^{-5}$  \\
Thermal conduction in the gas phase,$ k_z$ $(W/m.k)$& 0.09                  \\
Specific heat of gas, $C_{pg}$ $(J/mol.k)$                   & 30.7                  \\
Specific heat of adsorbed phase,  $C_{pa}$ $(J/mol.k)$       & 30.7                  \\
Radius of the pellets, $r_p$ $(m)$                           & $1 \times 10^{-3}$    \\
Specific heat capacity of the adsorbent, $C_{ps}$ $(J/kg.K)$ & 1070                  \\
Molar loading scaling factor, $q_s$ $(mol/kg)$               & 5.84                  \\
Mass transfer coefficient for $CO_2$, $k_{CO_2}$ $(1/s)$     & 0.1631                \\
Mass transfer coefficient for $N_2$, $k_{N_2}$ $(1/s)$       & 0.2044                \\
Purge Pressure, $P_L$ $(Pa)$                                 & $1 \times 10^{4}$     \\
Adsorption pressure, $P_0$ $(Pa)$                            & $1\times 10^{5}$      \\
Feed temperature of flue gas, $T_0$ $(K)$                   & 313.15                \\ \hline
\end{tabular}
\label{Tabela_parametros}
\end{table}

 To solve the system of partial differential equations, we begin by discretizing the spatial variables. In this process, we employ the finite volume method (FVM) and utilize an essentially non-oscillatory scheme known as WENO. Our discretization involves subdividing the equations into 10 volume elements. This value was defined based on a sensitivity analysis. Subsequently, we solve the resulting ordinary differential equations (ODEs) using MATLAB's rigid solver, ode15s. This discretization and equation resolution process ensures an effective and precise solution for the system of PDEs under study, allowing for a detailed analysis of its characteristic dynamic behavior.
This numerical solver was run on a computer featuring an AMD Ryzen 9 7950X processor, 128 GB of RAM, and an RTX 4090 GPU.

\begin{table}[]
\caption{PSA model boundary conditions}
\renewcommand{\arraystretch}{1.7} 
\begin{tabular}{>{\raggedright\arraybackslash}p{2cm}>{\centering\arraybackslash}p{2cm}>{\centering\arraybackslash}p{2.5cm}>{\centering\arraybackslash}p{2.5cm}>{\centering\arraybackslash}p{2.5cm}}
\hline
\multicolumn{3}{l}{Cycle step}                                                                                                    &  &  \\ \hline
\multicolumn{1}{c}{\multirow{2}{*}{1. Pressurization}} & \multicolumn{1}{c}{$Z = 0^+$} & \multicolumn{1}{c}{$P^* = P^*_L$} & \multicolumn{1}{c}{$y_i = y_{feed}$} & $T^* = 1$    \\
\multicolumn{1}{c}{}                                   & $Z = 1^-$                      & $\frac{\partial P^*}{\partial Z} = 0$                     & $\frac{\partial{y_i}}{\partial Z} = 0 $          & $\frac{\partial T^*}{\partial Z} = 0$              \\
\multirow{2}{*}{2. Adsorption}                         & \multicolumn{1}{c}{$Z = 0^+$} & \multicolumn{1}{c}{$P^* = P_high$} & \multicolumn{1}{c}{$y_i = y_{feed}$} & $T^* = 1$    \\
& $Z = 1^-$                      & $P^* = 1$                     & $\frac{\partial\bar{y_i}}{\partial Z} = 0 $                     & $\frac{\partial T^*}{\partial Z} = 0$   \\
\multirow{2}{*}{3. Heavy reflux}                              & \multicolumn{1}{c}{$Z = 0^+$} & \multicolumn{1}{c}{$P^* = P_high$} & \multicolumn{1}{c}{$y_i = y_{feed}$} & $T^* = 1$    \\
& $Z = 1^-$                      & $P^* = 1$                     & $\frac{\partial\bar{y_i}}{\partial Z} = 0 $                     & $\frac{\partial T^*}{\partial Z} = 0$   \\
\multirow{2}{*}{4. Depressurization}                           & \multicolumn{1}{c}{$Z = 0^+$} & $P^* = P^*_L$        &  $\frac{\partial{y_i}}{\partial Z} = 0 $                    & $\frac{\partial T^*}{\partial Z} = 0$    \\
     & $Z = 1^-$                     & $\frac{\partial P^*}{\partial Z} = 0$                      & $\frac{\partial{y_i}}{\partial Z} = 0 $                      & $\frac{\partial T^*}{\partial Z} = 0$    \\
\multirow{2}{*}{5. Light reflux}                              & \multicolumn{1}{c}{$Z = 0^+$} & $P^* = P^*_L$                     & $\frac{\partial{y_i}}{\partial Z} = 0 $                     &  $\frac{\partial T^*}{\partial Z} = 0$  \\
      & $Z = 1^-$                      & $P^* = P^*_L$          & $y_i = y_{i,ads}$                     &  $T^* = T^*_{ads}$  \\ \hline
\end{tabular}
\label{Contorno}
\end{table}

\subsection{Data Acquisition and Data Processing}\label{Data_Acquisition}

The phenomenological model of the PSA process played an essential role as a virtual plant for generating synthetic data. The controlled generation of synthetic data was done through the design of the experiments (DoE) allowing the exploration of various scenarios and operational conditions, enabling the surrogate model training in challenging situations that may be difficult to encounter in practice. This, in turn, leads to surrogate models that are robust and capable of handling a wider range of real-world scenarios, contributing to improved performance. To do so, we considered the PDE model as a virtual plant accessed through a software-in-the-loop (SIL).

Hence, the first step in obtaining and collecting synthetic data was the design of the experiments. The experimental conditions were defined based on a pseudo-random sampling technique called latin hypercube sampling (LHS) \citep{McKay}. LHS provides a solid statistical foundation for the experiments conducted in the virtual plant. This method combines random and stratified sampling techniques to generate sets of nearly random samples of various variables, avoiding unwanted correlations in the input variables of the DNN model.

The LHS's first step involves sampling each variable's cumulative probability density function. Subsequently, the variable samples are organized based on their correlations to prevent undesired correlations \citep{McKay}.
Therefore, the input variables of the surrogate model (in this case: $t_{press}$, pressurization step time; $t_{depress}$, depressurization step time; $t_{ads}$, adsorption step time; $t_{LH}$, light reflux step time; $t_{HR}$, heavy reflux step time) were defined using the latin hypercube sampling technique to be incorporated into the virtual plant and generate synthetic data regarding $CO_2$ purity ($Pur_{CO_2}$), $CO_2$ recovery ($Rec_{CO_2}$) in the cyclic steady-state (CSS). The minimum limits define the limits for each operational variable
$u_{min}$ = [10, 10, 10, 10, 10]s and maximum limits $u_{max}$ = [1000, 1000, 1000, 1000, 1000]s.

\begin{figure}[t]
\centering
\includegraphics[width=10 cm]{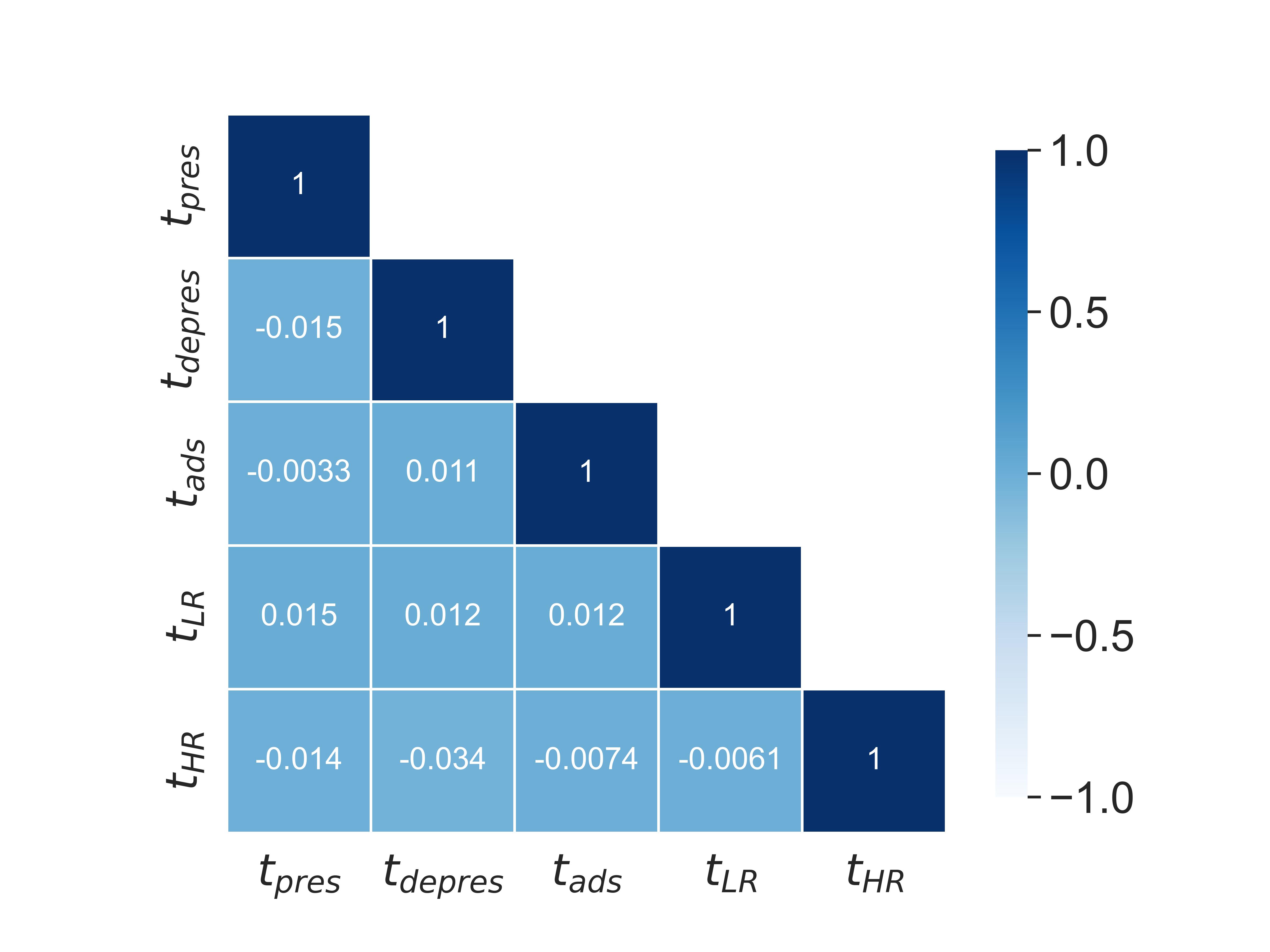}
\caption{Heatmap of correlations among LHS input signals.}
\label{Correlation}
\end{figure}

\begin{figure}
\centering
\includegraphics[width=14 cm]{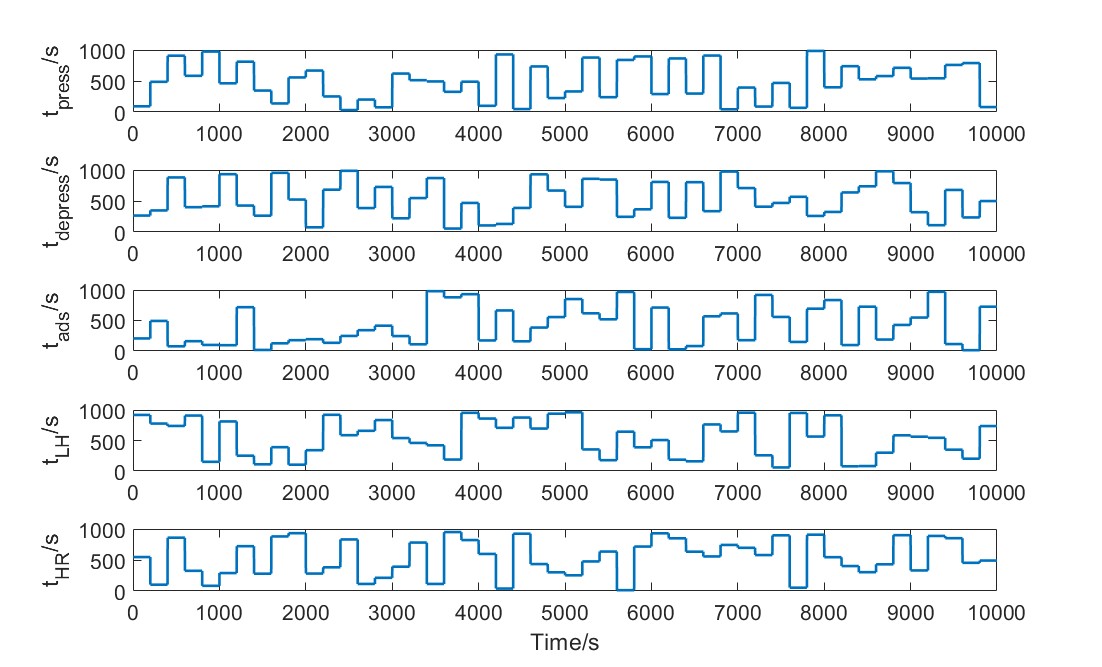}
\caption{Input representation for LHS-designed experiments for phenomenological model perturbation.}
\label{lhs}
\end{figure}

Following the generation of the input matrix using the latin hypercube sampling method, a correlation assessment among the variables was conducted. Figure \ref{Correlation} illustrates a heat map of all the model's input variables. The results reveal that the correlations exhibit values near zero, suggesting that the input space provided by LHS was thoughtfully designed. This effectively mitigates data distortions and, consequently, ensures that the training process remains free from unwanted biases.
The LHS was used to design 3,000 experiments. Afterward, these disturbances were introduced to the phenomenological model to produce datasets constituting the model's output matrix. Figure \ref{lhs} showcases the input representation to introduce disruptions in the phenomenological model. Each experiment was subsequently implemented in the PSA system until it achieved a cyclic steady state.

The data set obtained from the DoE was divided randomly into three subsets according to the following proportion: 75\% for training data, 15\% for validation data, and 15\% for test data.
This work structured the data using a multiple-input single-output (MISO) strategy. 

\unskip

\unskip

\subsection{DNN Identification}

After the data collection and preparation, the next step is the configuration of the non-linear function approximator, which in this case is the DNN model. DNN has several parameters to be estimated. One essential step before estimating these parameters is defining the model structure, the hyperparameters. These variables dictate the training and performance of a DNN model. The hyperparameters can be divided into two sets: model and algorithm parameters. The number of layers, neurons, activation function, and layer type are in the first set. In the second set are the tuning parameters of the optimization algorithm, which include the learning rate, momentum, learning rate decay, and several epochs for gradient-based optimizers. All those hyperparameters together form a space of both discrete and continuous variables. Due to their number and different nature, their proper estimation is a complex task. However, optimally adjusting these hyperparameters can enhance the network's generalization capability and mitigate the risk of overfitting.

The usual approach in the machine learning field is trial-and-error-based methods, such as random and grid search for hyperparameter tuning \citep{JMLR}. Due to the nature of these methods, they are computationally heavy and not efficient. On the other hand, the most recent in this field was made a few years ago by \cite{li2018hyperband}, where a method called HYPERBAND was proposed. This method was demonstrated to be significantly faster in search speed than random grid search while providing a statistical basis for a consistent search of the best set of parameters. This method is based on a formulation of the hyperparameter optimization problem as a pure-exploration problem. It uses a present resource allocated to perform random samples of the search space. The method performs a cuddled loop where an inner loop based on the SUCESSIVEHALVING method \citep{jamieson2015nonstochastic} is coupled to the outer loop based on a grid search. The application of the HYPERBAND requires the definition of the maximum number of epochs and the discarded proportion of non-ideal configurations in each. The HYPERBAND algorithm is one of the most advanced strategies in the field.

Before applying the HYPERBAND algorithm for optimization, the initial step involves defining the hyperparameters of interest and their corresponding search spaces. These hyperparameters may encompass the learning rate, the number of layers in the neural network, batch size, and other factors influencing the model's performance. Each hyperparameter is bounded by a specific search space, typically representing a range of possible values for that parameter. By defining these search spaces, the HYPERBAND algorithm can explore various combinations of hyperparameters to find the optimal set that maximizes the model's performance.
In this study, the initial learning rate, the number of dense layers, the activation functions in each layer, and the number of neurons in each layer were selected as the focus hyperparameters in the quest for the ideal set of parameters for the model. In this study, Table \ref{TAB:Hyperparameters} describes the hyperparameter search space.

\begin{table}[H]
\renewcommand{\arraystretch}{1.5} 
\caption{Hyperparameter search space for HYPERBAND and best hyperparameters for each performance indicator for DNN.\label{tab2}}
\begin{tabularx}{\linewidth}{m{1.5cm}m{5.5cm}CC}
    \toprule
    \multicolumn{4}{c}{\textbf{Performance parameter models}} \\
    \midrule
    \textbf{Models} &\textbf{Hyperparameters} & \textbf{Search space} & \textbf{Best hyperparameters}\\
    \midrule
    \multirow{6}{*}{\rotatebox[origin=c]{90}{\centering{$CO_2$ purity}}} & Initial learning rate & $1\times{10}^{-4}$,$1\times{10}^{-3}$, $1\times{10}^{-2}$ & $1\times{10}^{-3}$ \\
    & Number of dense layers & \{1,2,3,4,5,6\} & 5 \\
    & Activation function in the dense layers & \{relu, tanh\} & \{relu, relu, relu, relu, linear\} \\
    & Number of neurons in the dense layer & from 50 to 300 with 10 step & \{290, 100, 280, 200, 1\} \\
    & Number of parameters for the layers & -- & \{1740, 29100, 28280, 56200, 201\} \\
    \midrule
    \multirow{6}{*}{\rotatebox[origin=c]{90}{\centering{$CO_2$ recovery}}} & Initial learning rate & $1\times{10}^{-4}$,$1\times{10}^{-3}$, $1\times{10}^{-2}$ & $1\times{10}^{-3}$ \\
    & Number of dense layers & \{1,2,3,4,5,6\} & 4 \\
    & Activation function in the dense layers & \{relu, tanh\} & \{ relu, relu, relu, linear\} \\
    & Number of neurons in the dense layer & from 50 to 300 with 10 step & \{220, 60, 80, 1\} \\
    & Number of parameters for the layers & -- & \{1320, 13260, 4880, 81\} \\
    \bottomrule
\end{tabularx}
\label{TAB:Hyperparameters}
\end{table}

The HYPERBAND algorithm iteratively adjusts the model's hyperparameters throughout the optimization process to enhance its performance for the defined objective function. This objective function is evaluated using the training and validation data with the hyperparameters selected within the specified search space. In this context, the function encompasses a loss function that assesses the model's performance on the training and validation datasets.

\subsection{DNNs Training and Testing}

\begin{figure}[h!]
	\centering
    \begin{subfigure}{0.45\textwidth}
        \centering
		\includegraphics[scale=.25]{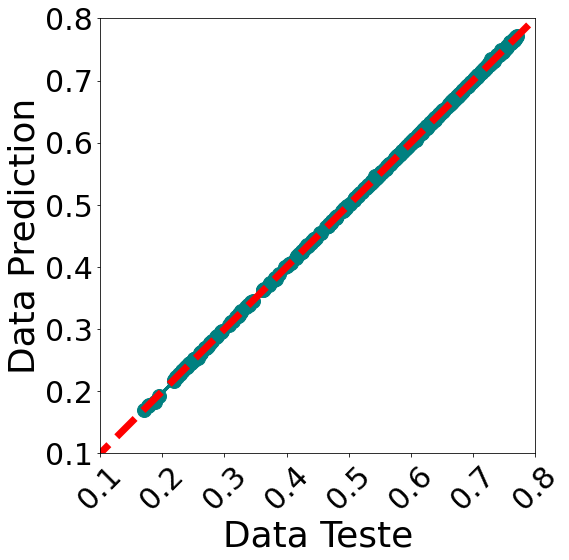}
	\caption{$CO_2$ Purity}
	\label{FIG:Paridade_purity}
\end{subfigure}
    \centering
\begin{subfigure}{0.45\textwidth}
	\centering
		\includegraphics[scale=.25]{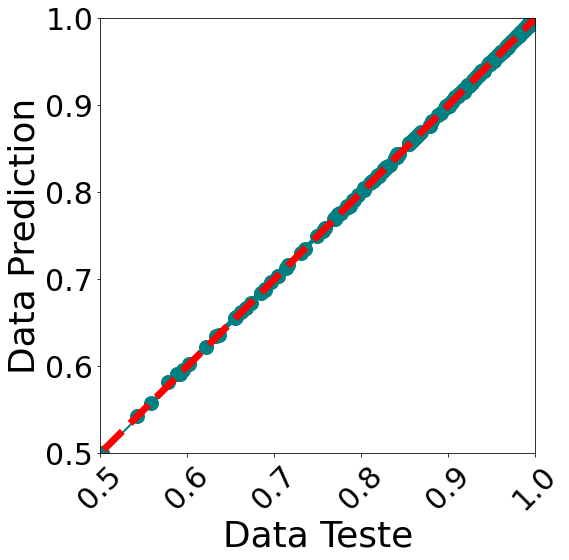}
	\caption{$CO_2$ Recovery}
	\label{FIG:Paridade_recovery}
\end{subfigure}
    \caption{Parity graphics of the model test.}
    \label{FIG:Paridade}
\end{figure}

With the selection of DNN structures, the subsequent phase involves estimating and testing the models' internal parameters. The ADAM algorithm \citep{kingma2017adam} was used to train the models based on the configuration given in Table \ref{TAB:Hyperparameters}. Following the training process, a testing step is performed using a set of experiments that were not part of the model training process. The testing results are illustrated in Figure \ref{FIG:Paridade}, enabling a visual assessment of the fitness of each model.
More specifically, the parity plot visually compares model predictions and actual test data. This chart assesses the model's performance and ability to capture system behavior accurately. The random distribution of points along the graph's diagonal line highlights the model's accuracy. The parity plot demonstrates that the AI model's predictions consistently align with the test data across the entire test range.

Table \ref{TAB:metricas} presents the AI model's performance results. To assess the models' performance, we employed standard metrics such as mean absolute error (MAE) and mean squared error (MSE), widely recognized for measuring the discrepancy between actual values and predictions \citep{Qi_2020}. Furthermore, the identification and statistical analysis of a model's predictions can be important in tracking the potential limitations of the model and assessing the need to readjust its parameters.

The results reveal that the models exhibit significantly low values for both MAE and MSE concerning the test data. This attests to the successful model identification and their ability to make accurate predictions. These low error levels underscore the models' reliability and capacity to capture the PSA model's performance metrics. Based on the results in Table \ref{TAB:metricas}, the obtained error is acceptable and does not justify the increased computational effort.

\begin{table}[]
\renewcommand{\arraystretch}{1.5} 
\caption{Final MAE and MSE to test dataset.}
\label{tbl:MAE_MSE_Test_Dataset}
\begin{tabular}{cccc}
\cline{1-3}
\multicolumn{1}{l}{Variable} & MSE                           & MAE                         &  \\ \cline{1-3}
$CO_2$ Purity                 & $1.13 \cdot 10^{-6}$   & $7.44 \cdot 10^{-4}$ &  \\
$CO_2$ Recovery               & $1.14 \cdot 10^{-6}$ & $6.98 \cdot 10^{-4}$  &  \\ \cline{1-3}
\label{TAB:metricas}
\end{tabular}
\end{table}

\subsection{Multi-objective Optimization Problem for PSA unit}

This section presents the mathematical design of the optimization problem proposed here. A multi-objective strategy was followed as $CO_2$ capture in a PSA unit must accommodate the trade-off between unit performance parameters. Instead of exclusively focusing on a single metric, multi-objective optimization considers both objectives simultaneously. This assists in finding solutions that represent the best compromise between purity and recovery, considering the specific constraints and requirements of the process.
Well-adjusted step times can significantly impact the final product's purity and the desired component's recovery. A longer time in the adsorption step may increase purity but potentially reduce recovery. Conversely, a shorter time may result in higher recovery but reduced purity. Therefore, optimizing step times is essential to achieve the desired balance between purity and recovery, tailoring the process to specific application needs. In this work, step times were used as decision variables in the optimization problem.
All process outputs were evaluated in cyclic steady state (CSS) so that the previously mentioned constraints are end-point constraints. The inputs, decision variables $u$, are the same inputs from the data-based model; they were subjected constraints
$[u_{min},u_{max}]$ imposed along the trajectory until the CSS is reached.

In mathematical terms, the optimization problem at hand can be calculated as per the following problem:

\begin{equation}
 \begin{cases}
     &max \, \, \text{Pur}_{CO_2}(u, \beta),  \\ 
     &max \, \,  \text{Rec}_{CO_2}(u, \beta),
    \end{cases}
    \label{funcao_obj}
\end{equation}

subject to:

\begin{align}
& u \in [u_{\text{min}}, u_{\text{max}}] \\
& 0\% \leqslant \text{Pur}_{CO_2}(u, \beta), \text{Rec}_{CO_2}(u, \beta) \leqslant 100\%
\label{restricoes}
\end{align}

where:

\begin{equation}
Pur_{CO_2}\ =\frac{{moles}_{CO_2,\left.out\right|Depres}+{moles}_{CO_2,\left.out\right|LR}}{{moles}_{total,\left.out\right|Depres}+{moles}_{total,\left.out\right|LR}}
    \label{pureza}
\end{equation}

\begin{equation}
Rec_{CO_2}\ =1-\frac{{moles}_{CO_2,\left.out\right|Ads}}{{moles}_{CO_2,\left.in\right|Pres}+{moles}_{CO_2,\left.in\right|Ads}}
    \label{recuperacao}
\end{equation}

Equations \ref{moles_1} - \ref{moles_7} are the necessary equations referring to the number of moles at different steps to calculate purity and recovery
of $CO_2$ over the cycles \citep{doi:10.1021/acs.iecr.9b02383}.

\begin{equation}
{moles}_{CO_2,\left.out\right|Depres}=\frac{P_0L}{RT_0}\varepsilon A\int_{0}^{\tau_{Depres}}{{\bar{v}}_Zy_{CO_2}\frac{P^*}{T^*}d\tau\ \ \ \ \forall Z=0^+}
    \label{moles_1}
\end{equation}

\begin{equation}
{moles}_{CO_2,\left.out\right|LR}=\frac{P_0L}{RT_0}\varepsilon A\int_{0}^{\tau_{LR}}{{\bar{v}}_Zy_{CO_2}\frac{P^*}{T^*}d\tau\ \ \ \ \forall Z=0^+}
    \label{moles_2}
\end{equation}

\begin{equation}
{moles}_{total,\left.out\right|Depres}=\frac{P_0L}{RT_0}\varepsilon A\int_{0}^{\tau_{Depres}}{{\bar{v}}_Z\frac{P^*}{T^*}d\tau\ \ \ \ \forall Z=0^+}
    \label{moles_3}
\end{equation}

\begin{equation}
{moles}_{total,\left.out\right|LR}=\frac{P_0L}{RT_0}\varepsilon A\int_{0}^{\tau_{LR}}{{\bar{v}}_Z\frac{P^*}{T^*}d\tau\ \ \ \ \forall Z=0^+}
    \label{moles_4}
\end{equation}

\begin{equation}
{moles}_{CO_2,\left.out\right|Ads}=\frac{P_0L}{RT_0}\varepsilon A\int_{0}^{\tau_{Ads}}{{\bar{v}}_Zy_{CO_2}\frac{P^*}{T^*}d\tau\ \ \ \ \forall Z=1^-}
    \label{moles_5}
\end{equation}

\begin{equation}
{moles}_{CO_2,\left.in\right|Pres}=\frac{P_0L}{RT_0}\varepsilon A\int_{0}^{\tau_{Pres}}{{\bar{v}}_Zy_{CO_2}\frac{P^*}{T^*}d\tau\ \ \ \ \forall Z=0^+}
    \label{moles_6}
\end{equation}

\begin{equation}
{moles}_{CO_2,\left.in\right|Ads}=\frac{P_0L}{RT_0}\varepsilon A\int_{0}^{\tau_{Ads}}{{\bar{v}}_Zy_{CO_2}\frac{P^*}{T^*}d\tau\ \ \ \ \forall Z=0^+}
    \label{moles_7}
\end{equation}

To solve the optimization problem outlined in Equation \ref{funcao_obj} - \ref{recuperacao}, we applied a constrained sliding particle optimization algorithm, as proposed by \cite{math9151808}. This algorithm has proven to be a practical choice for optimizing cyclic adsorption processes.

To develop a solid assessment and gain important insights into the decision-making process, the $CO_2$ productivity and the energy required for the CCS cycle were calculated, as shown in the following equations.

\begin{equation}
    CO_2 \; productivity = \frac{mol\; CO_2 \; in \; the \; product}{mass \; of \; adsorbent \times cycle \; time} 
    \label{productivity}
\end{equation}

\begin{equation}
    Energy \; required = \frac{energy \; required \; for \; all \; steps }{mass  \; of \; CO_2 \; collected \; in the \; product \; per \; cycle}
\end{equation}

Specifically in terms of the optimization algorithm, this research implemented a variant of PSO, the constraints-based sliding particle swarm optimizer (CSPSO), as proposed by \cite{math9151808}, a metaheuristic algorithm grounded on population principles and multilocal minima search. The selection of CSPSO was driven by its ability to handle local minima and to draw the process feasible operating regions. Within the CSPSO algorithm, particles, constituted by decision variables, navigate the problem’s search space. Each particle’s position signifies a potential solution correlated with the set of decision variables. Consequently, PSO utilizes elementary mathematical operations to direct each particle towards areas of promise within the search space. This progression considers the optimal positions found by the individual particle - termed as the individual component, and the most favorable value discovered across all particles, termed as the social component. With each iteration, a new set of positions is obtained. The optimization procedure within PSO persists until a termination condition, such as achieving a maximum number of
iterations or obtaining a satisfactory solution, is met. Upon satisfaction of the termination condition, the algorithm concludes, delivering the optimal solution found during the optimization process.

\subsection{Feasible Operation Region Map}

In this study, we repurposed the results obtained from assessing the objective function during the optimization problem-solving process to construct a feasible operating region of the process closely aligned with ideal conditions. To build the FOR, we utilized the extended Fisher-Snedecor statistical test. This test was employed in the context of the multi-objective optimization problem addressed in this research to outline the region near the Pareto front.
The Equation \ref{Fisher} shows the Fisher-Snedecor test applied to this study, with 95\% confidence level ($\alpha$) \citep{REBELLO2022590,math9243152}.

\begin{equation}
    L_i (\boldsymbol{\theta}_k,\boldsymbol{\lambda_k}) \leq L_i (\boldsymbol{\theta}_j^*,\boldsymbol{\lambda_j^*} )+ \frac{(n_k n_j)}{(n_k-n_{\boldsymbol{\theta}}+n_i )}  F_\alpha (n_j,n_k-n_{\boldsymbol{\theta}}+1)
    \label{Fisher}
\end{equation}

\begin{figure}[b]
\centering
\includegraphics[width=8 cm]{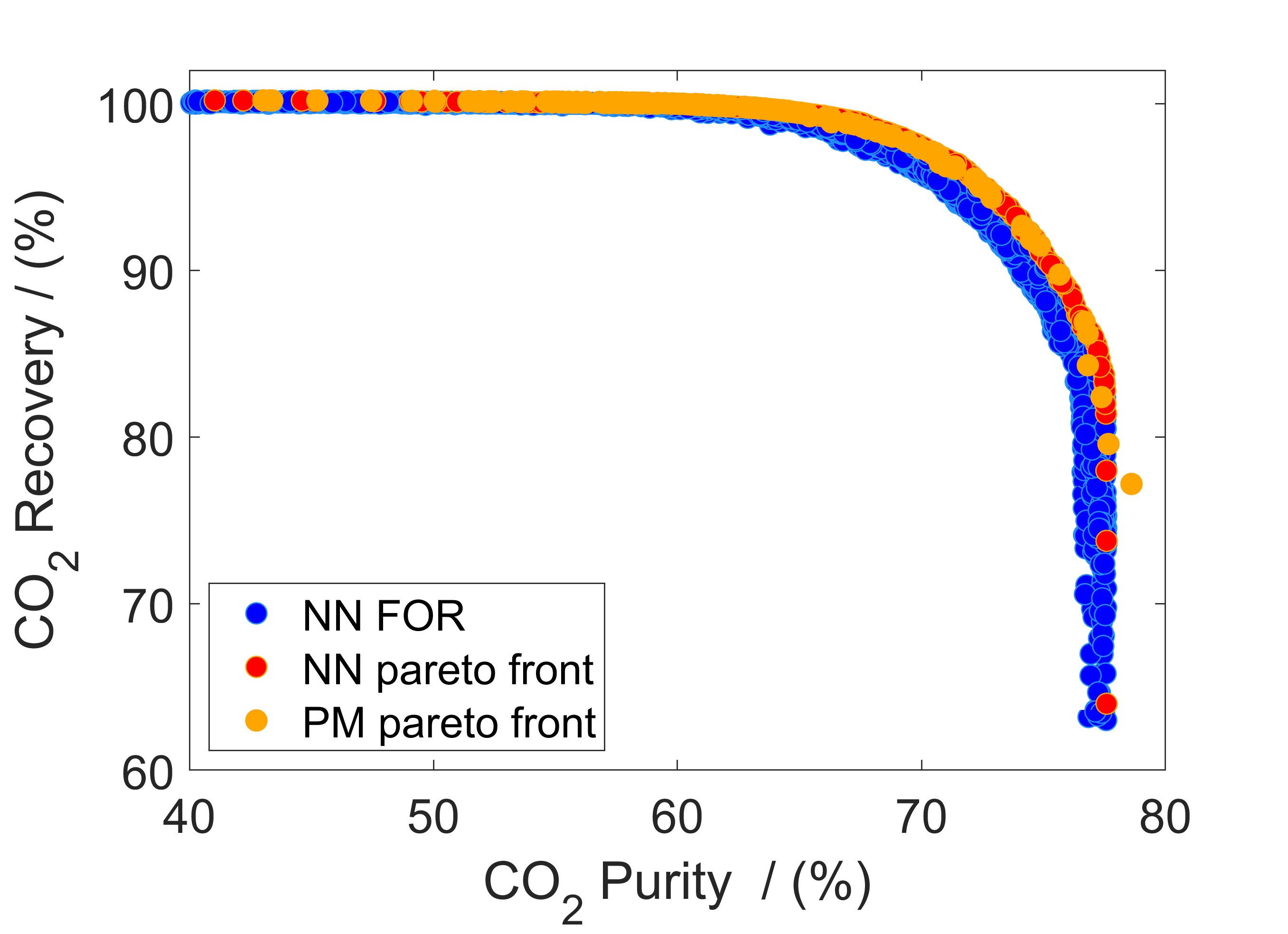}
\caption{Overlap of the Pareto optimal front for the DNN model and phenomenological model and feasible operation region defined by the Fisher-Snedecor test.}
\label{Fronteira_Pareto}
\end{figure}

In this context, the CSPSO optimizer was used with 50 iterations and 500 particles. Figure \ref{Fronteira_Pareto} illustrates the solution of the optimization problem using the data-driven model. The Pareto front obtained comprises 660 points, presenting significant diversity within the set of solutions. 

The optimality of the surrogate-based optimization was evaluated by introducing the decision variables corresponding to the 660 points from the Pareto front in the rigorous model. Hence, if there is an overlap between the rigorous model outputs and the Pareto front found in the surrogate-based optimization, we can consider it a reliable result.

This overlap guarantees the good results achieved in the optimization conducted through the substitute model based on the identified DNN. It was possible to use the optimal points obtained by the DNN model as input to the phenomenological model to establish a comparison with the substitute model. Consulting the phenomenological model takes approximately 600 seconds while using the empirical model requires only 0.1 seconds. This disparity in computational effort between both strategies justifies the option for the substitute approach. On the other hand, we are reducing the phenomenological optimization requirement to an on-demand evaluation of the rigorous model.
For instance, when using 50 particles and conducting 50 iterations, the DNN-based optimization shows convergence in 250 seconds, while the phenomenological model could take 259200 seconds, about 1000X more than the surrogate-based optimization. This notable discrepancy highlights the computational advantage of the DNN model. However, as presented here, this advantage must be used with caution. Therefore, we proposed to sample only the Pareto front of the surrogate-based optimization; this extra step adds 68429 seconds to the optimization procedure, which is still 273 times lower than the phenomenological-based optimization.

With the optimality verified, we can proceed to explore the surrogate-based results. In this work, we propose dividing the FOR into three groups using clustering techniques. Each subgroup will represent a zone on the Pareto front where each conflicting objective is favored. region 1 prioritizes one objective, region 3 prioritizes another, whereas region 2 represents a balance between the two. Furthermore, the operation region clusters provide flexibility for the process operation, thus constituting a tool capable of assisting in decision-making based on the Pareto map clusters. This tool can play an important role in process monitoring, providing relevant information about the behavior of the system's optimal conditions.

Therefore, the clustering method aggregates the population points within the FOR according to a metric. Among various possible metrics, the weighted average distance (WPGMA) was used, in which the (Euclidean) distance between the clusters is calculated as a simple average. The method aims to construct subgroups of data based on defined centroids, which will form each of the three regions. This is done by calculating the distance between the centroid and the surrounding points. The points that compose each region are those with the smallest distance from the corresponding centroid.

The next section will delve into the insights, comparisons, and knowledge gained regarding the system through the process operation map.

\section{Discussion}

One of the objectives of this work is to use the operation regions defined based on the methodology presented in this study to discuss the process behavior in the context of the resulting operating map.
Using the operational region derived from the Pareto front facilitates the process operating in terms of guiding the resource allocation. Hence, these results are produced as a decision-making support tool. 

In the optimization problem presented, there are two objectives, and the objective function is continuous. From the decision-maker's perspective, there are three options: an action that prioritizes one objective over the other, the opposite action, or an action that would strike a balance between the two objectives. Thus, to represent these possible courses of action, three clusters were used.

\begin{figure}[t]
\centering
\includegraphics[width=8 cm]{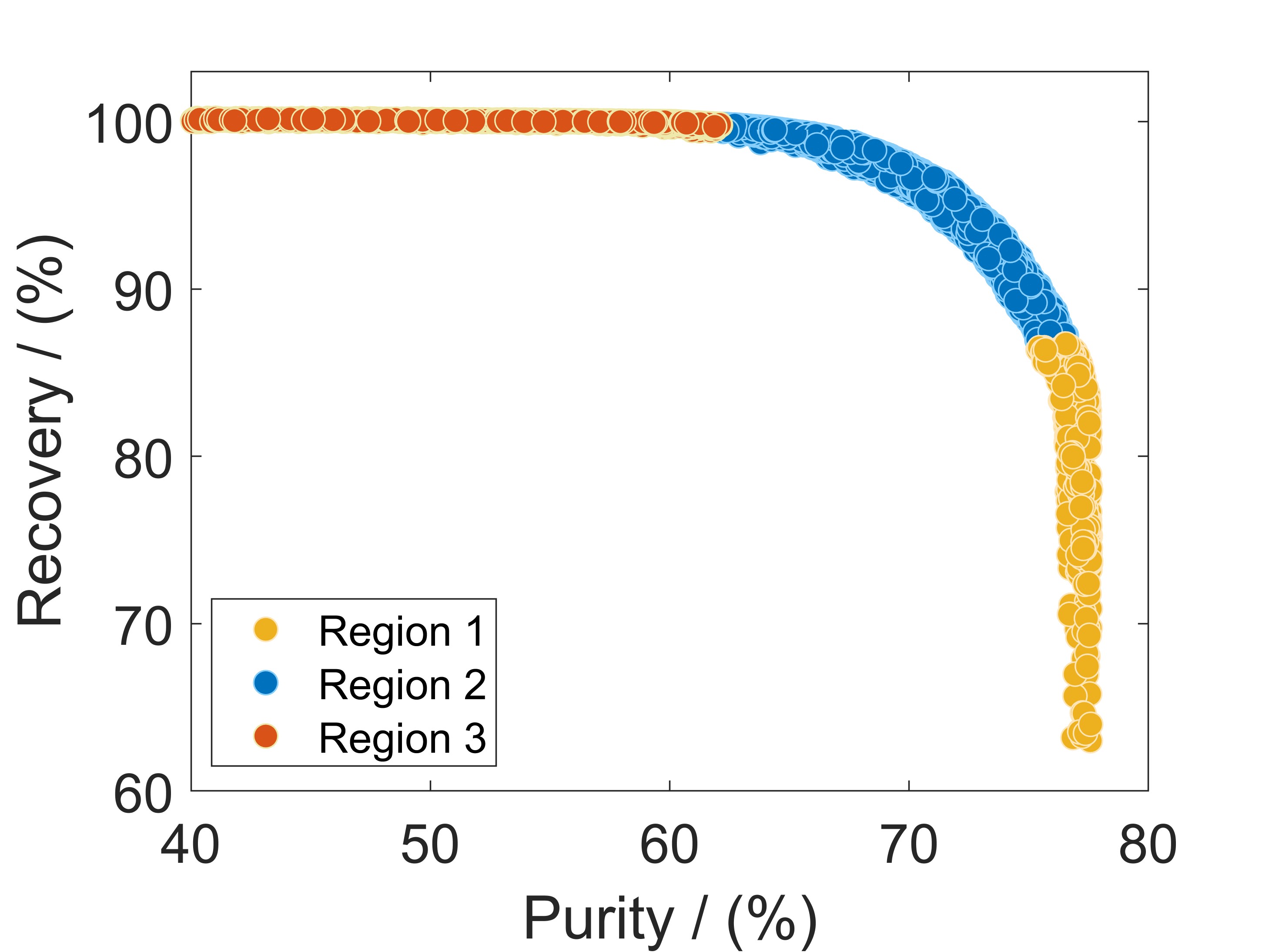}
\caption{Feasible operation regions defined by clustering.}
\label{Clusterization}
\end{figure}

The FOR was subdivided into three distinct operational regions presented in Figure \ref{Clusterization}, each with a specific focus and priorities. The first of these zones, denoted as region 1, primarily emphasizes maximizing $CO_2$ purity, striving to achieve the highest possible levels of purity within the constraints of the process. In region 2, the objective shifts to striking a delicate balance between $CO_2$ purity and $CO_2$ recovery, where both factors are optimized concurrently to meet specific operational requirements. Finally, region 3 prioritizes maximizing $CO_2$ recovery without considering the $CO_2$ purity. The division of FOR into these operational zones offers a more nuanced approach to process optimization, allowing for tailored solutions that align with the specific objectives and constraints of the system. Hence, the results are converted into a guideline for decision-making.

The operation map is also conducted from the perspective of decision variables associated with each respective cluster. This aspect is crucial as it allows for a more comprehensive evaluation of how the process can be effectively operationalized to achieve the desired objectives.

\begin{figure}[t]
	\centering
    \begin{subfigure}{0.45\textwidth}
        \centering
		\includegraphics[scale=.07]{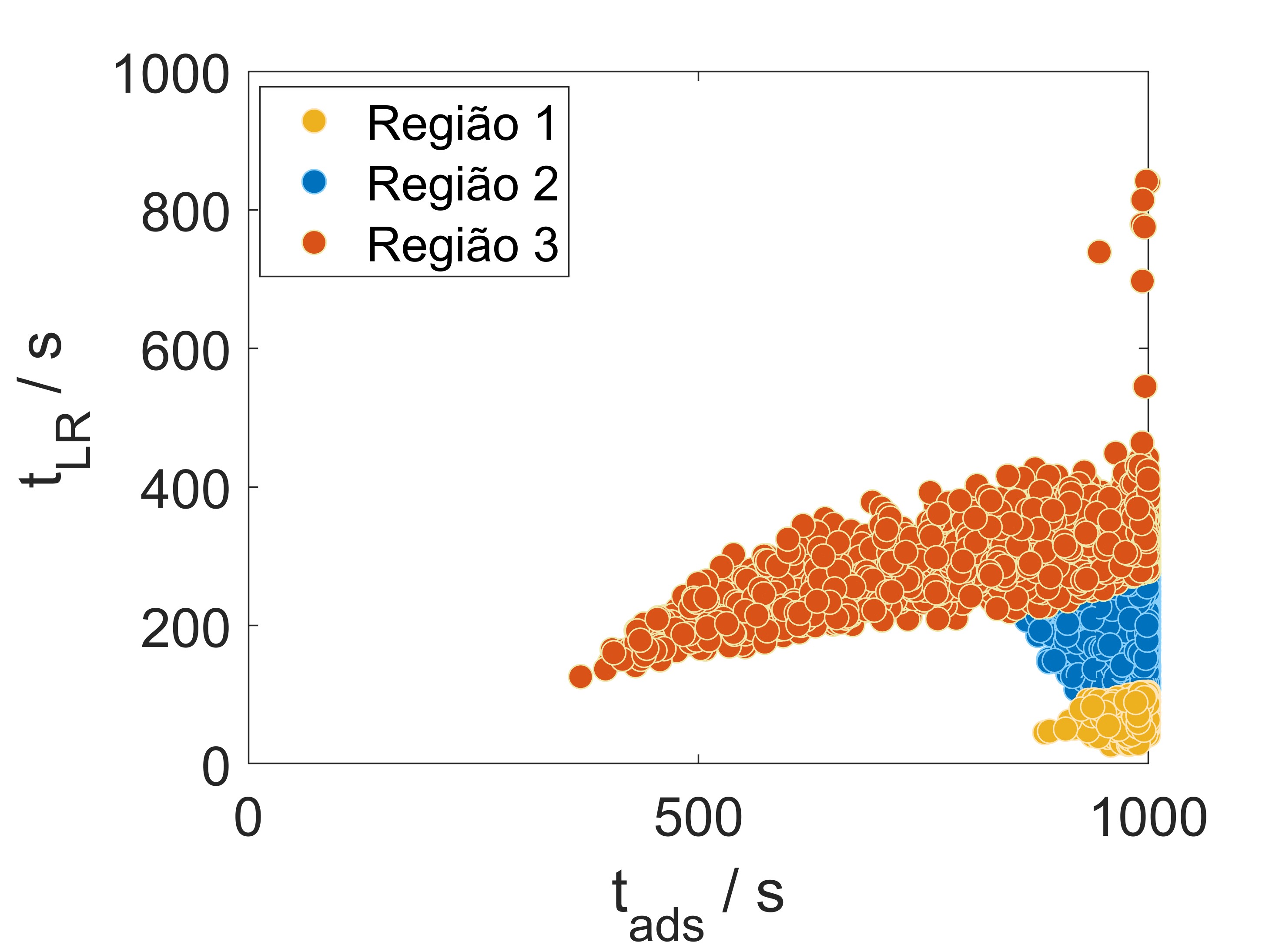}
\end{subfigure}
    \centering
    \begin{subfigure}{0.45\textwidth}
        \centering
		\includegraphics[scale=.07]{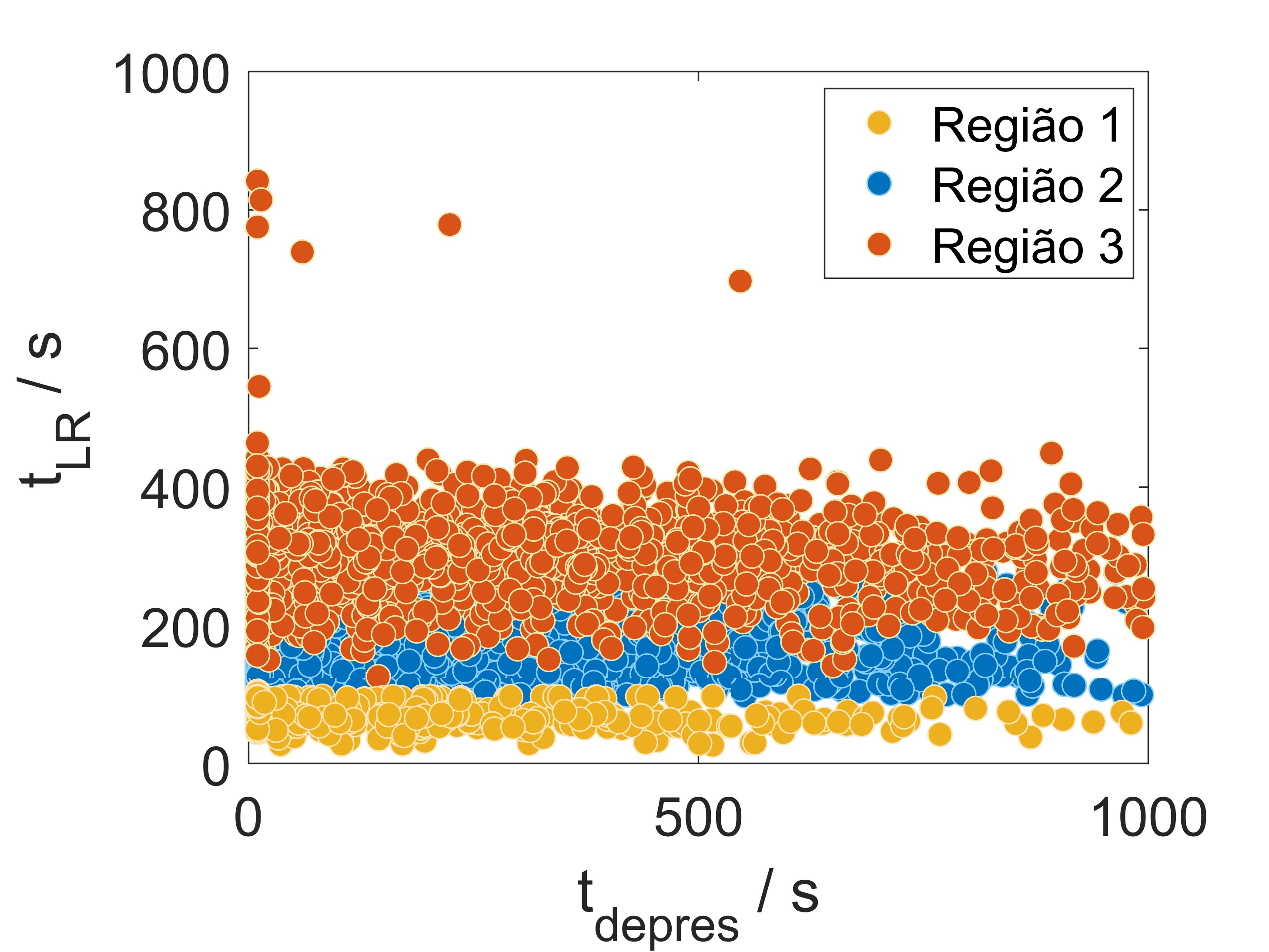}
\end{subfigure}
    \centering
\begin{subfigure}{0.45\textwidth}
	\centering
		\includegraphics[scale=.07]{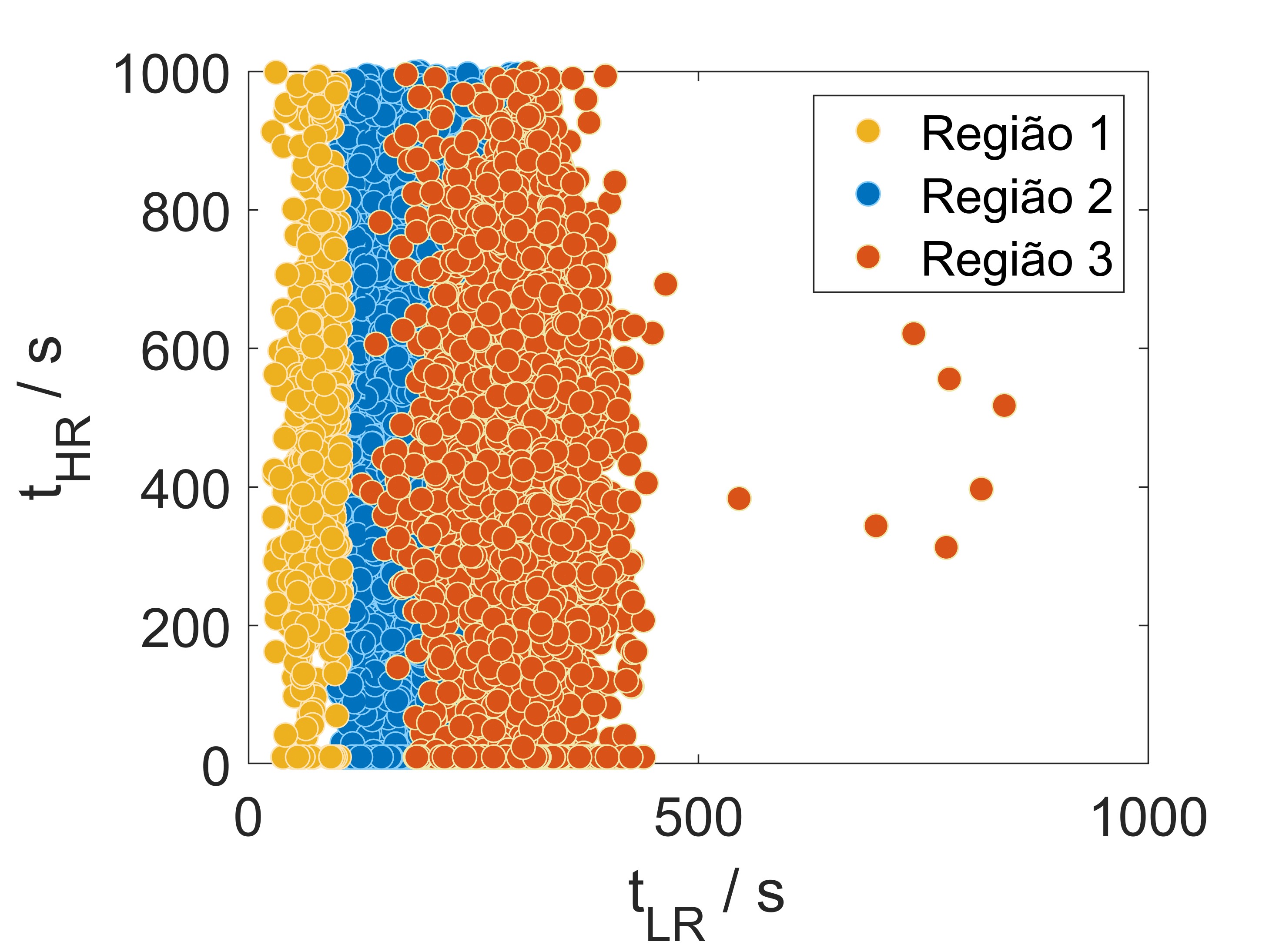}
  \end{subfigure}
      \begin{subfigure}{0.45\textwidth}
        \centering
		\includegraphics[scale=.07]{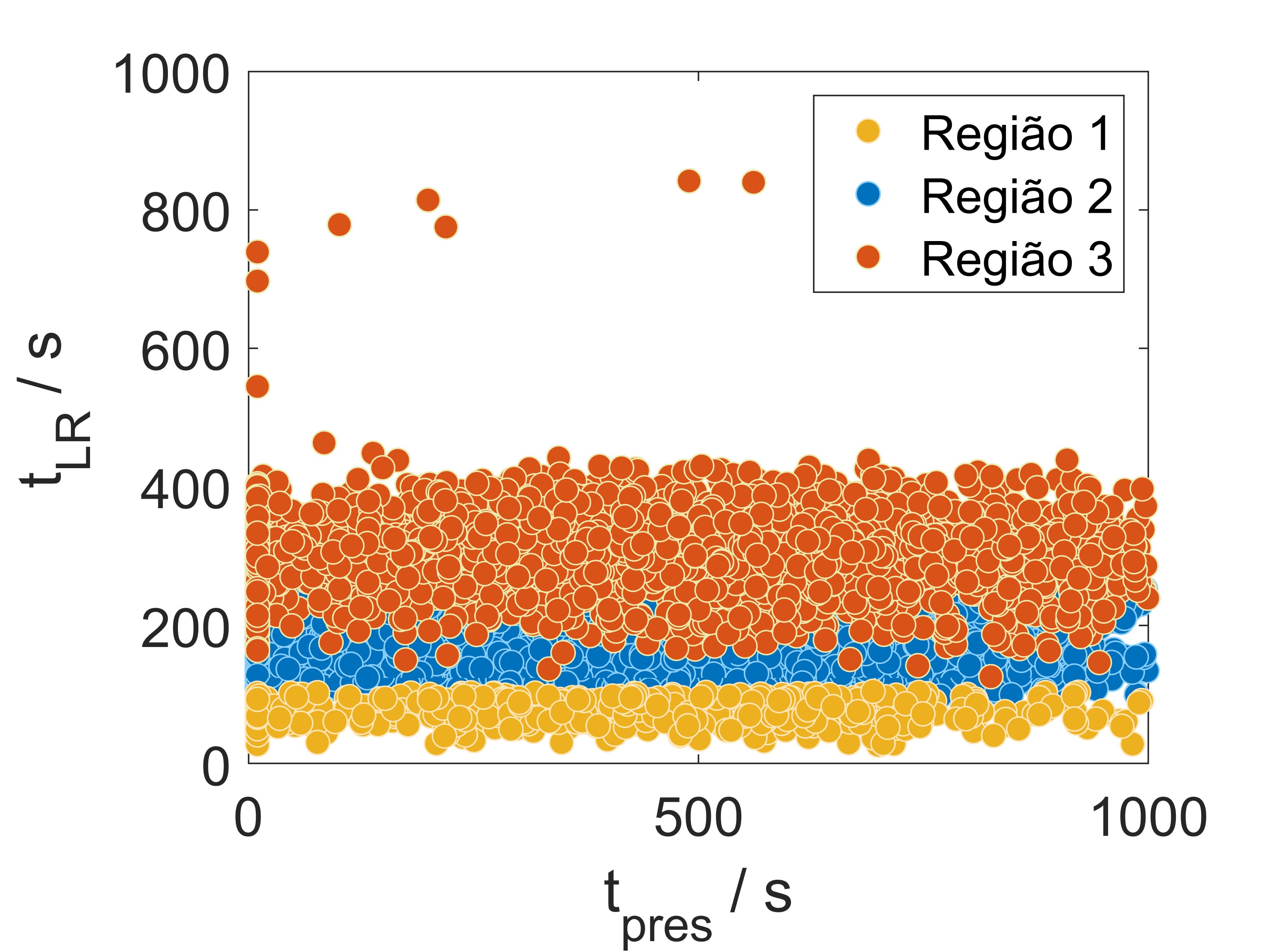}
\end{subfigure}
    \caption{Feasible operation map from the point of view of decision variables.}
    \label{FIG:map_decision_variable}
\end{figure}

Figure \ref{FIG:map_decision_variable} depicts the combinations of decision variables, their interrelationships, and how regions 1, 2, and 3 are associated with the operating variables.

It is important to note that in region 1, where $CO_2$ purity is the primary focus, the duration of the light reflux step is limited to around 100 seconds. This implies the process’s sensitivity to longer LR times, which could introduce impurities into the light reflux output stream and, consequently, reduce the purity of $CO_2$. In regions 2 and 3, the duration of the light reflux step increases, leading to a gradual decrease in $CO_2$ purity.

Furthermore, the interplay between LR time and $CO_2$ purity significantly affects process efficiency. In region 1, the LR time duration given by our methodology ensures high $CO_2$ purity but may limit the overall $CO_2$ recovery. In contrast, Regions 2 and 3 allow longer LR times, prioritizing higher $CO_2$ recovery but at the expense of slightly lower $CO_2$ purity. This delicate balance between purity and recovery is a crucial factor in optimizing the process and tailoring it to specific application requirements.

Assessing $CO_2$ purity from the standpoint of the adsorption step, it is evident that higher $CO_2$ purities require extended durations in this phase to ensure efficient $CO_2$ capture and reduce the presence of $N_2$ in the bed. Thus, during the depressurization and light reflux steps, the stream is enriched in $CO_2$.
On the other hand, $CO_2$ recovery is affected by the extended duration of the adsorption step. The longer $CO_2$ contaminates the output stream of the adsorption step, the lower the $CO_2$ recovery will be. These relationships become evident when revisiting Equations \ref{pureza} and \ref{recuperacao}.
The balance between purity and recovery is a critical factor in optimizing the PSA process. Achieving the desired purity while maximizing $CO_2$ recovery requires carefully adjusting the cycle durations.

For the remaining Press, Depres, and HR steps, a wide range of combinations is possible within a broad time frame, ranging from 10 to 1000 seconds. This suggests a sloppiness in these variables regarding the optimal conditions. 

On the other hand, the variability within this wide time range offers process engineers the flexibility to adapt the PSA process to different operating conditions and specific objectives. However, the emphasis on the correlation and constraints of $t_{LR}$ and $t_{ads}$ underscores the pivotal role of these parameters in determining the overall performance and trade-offs in the PSA process.

\begin{figure}
\centering
\includegraphics[width=14 cm]{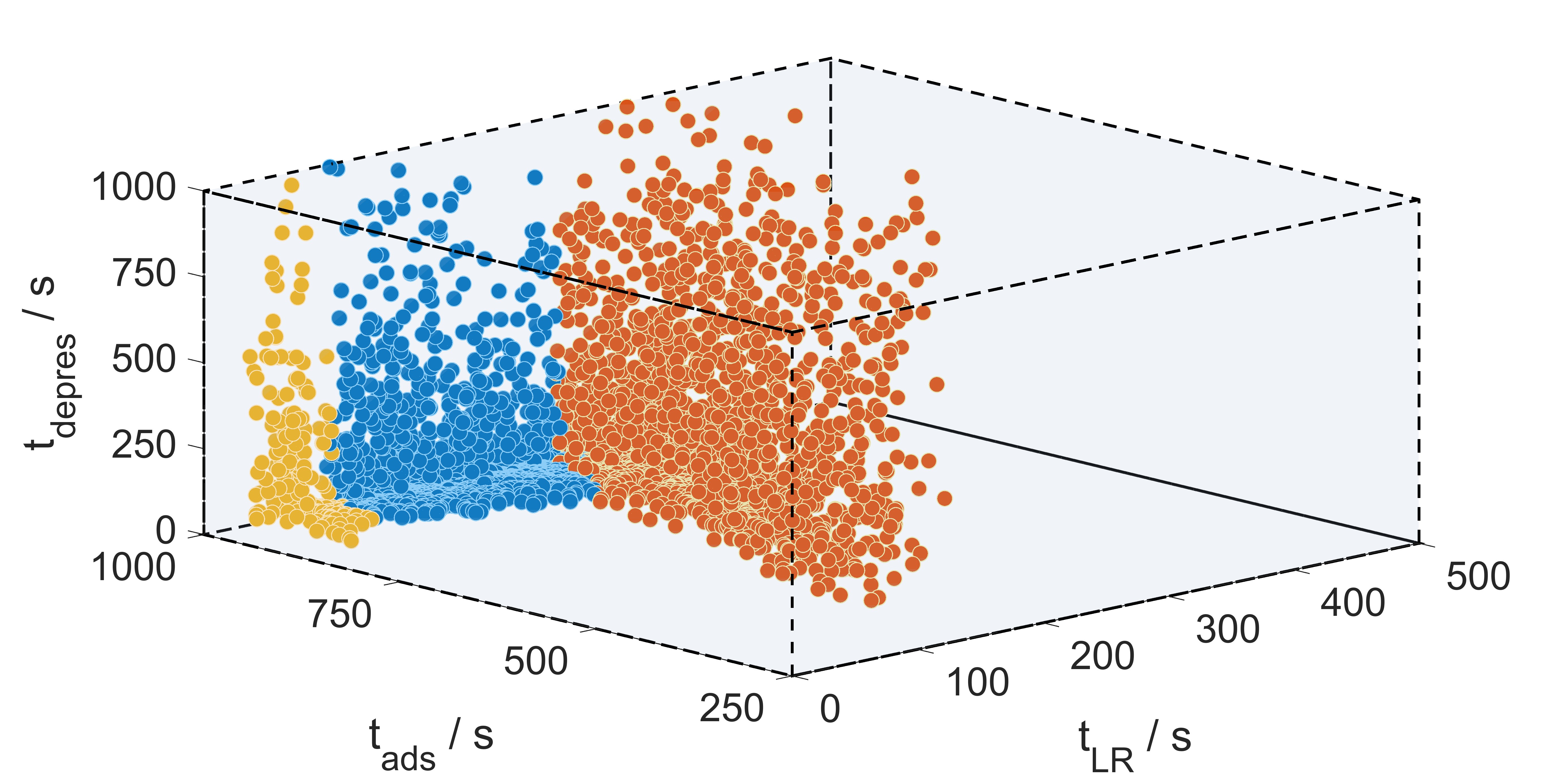}
\caption{Feasible operation map in 3D for $t_{depres}$, $t_{ads}$ and $t_{LR}$, (region 1 - yellow, region 2 - blue, region 3 - red)}
\label{3D_decision_variabels}
\end{figure}

A three-dimensional (3D) analysis, as represented in Figure \ref{3D_decision_variabels}, offers another perspective when evaluating decision variables. This methodology allows observation of particle dispersion, highlighting the interconnection of three operational variables within distinct clusters. The duration of the process steps appears to be a key factor that influences the purity and recovery of $CO_2$.

In this context, a sharp contrast is observed in the dispersion of data points in different regions. Region 1, which prioritizes $CO_2$ purity, presents less dispersion for the three variables. In contrast, region 3, which emphasizes the recovery of $CO_2$, shows a significantly higher degree of dispersion, suggesting a wider range of variable interactions.
This distinction highlights the specificity and precision required in adjusting operational variables to achieve high purity levels.

\begin{figure}
\centering
\includegraphics[width=15cm]{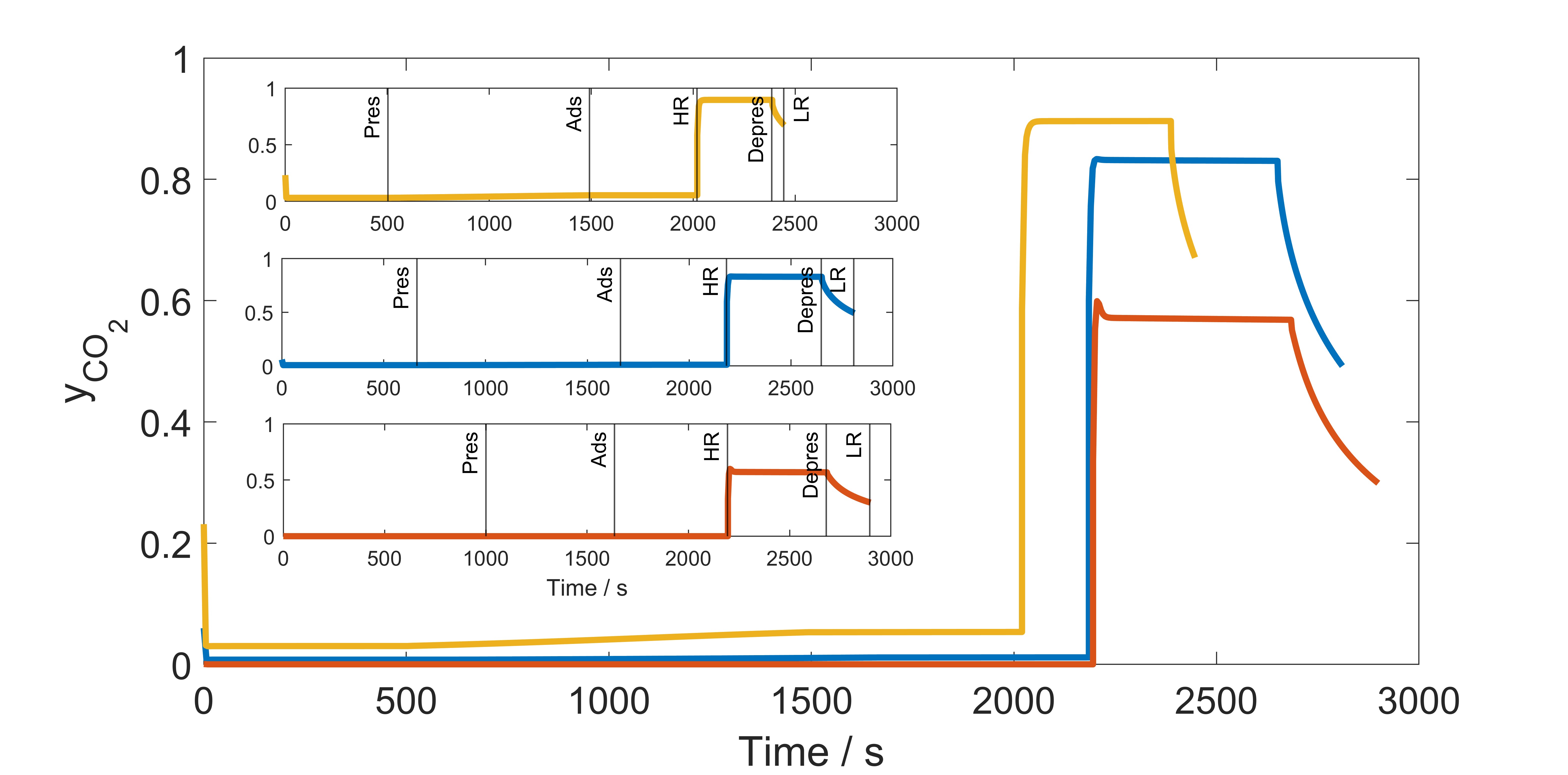}
\caption{History of $CO_2$ mole fractions for a cycle in the CSS for each selected point in each operating region. (region 1 - yellow, region 2 - blue, region 3 - red).\label{Dinamica}}
\end{figure}

To enrich our analysis, we present in Figure \ref{Dinamica} a representation of the system's dynamics in the CSS. As shown in Table \ref{tab:Mapa}, we selected a point in each operation region within the operations map, allowing us to evaluate the dynamic behavior at the end of each step. Table \ref{tab:Mapa} shows the corresponding performances for each point in the operating regions.

This operational profile suggests that while region 1 prioritizes $CO_2$ purity over recovery with moderate energy consumption, region 2 prioritizes $CO_2$ recovery with slightly better productivity. region 3 maximizes $CO_2$ recovery at the expense of higher energy demand and lower purity. The selection of the optimal operation region will depend on the specific objectives of the process and cost-benefit considerations related to energy consumption and the desired product quality.

\begin{table}[H]
\centering
\renewcommand{\arraystretch}{1.4} 
\setlength{\tabcolsep}{14pt} 
\begin{tabular}{lccc}
\hline
\textbf{Operational parameters} & \textbf{Region 1} & \textbf{Region 2} & \textbf{Region 3} \\ \hline
Time of pressurization step / s & 503.6 & 663.1 & 1001 \\
Time of depressurization step / s & 367.5 & 465.7 & 488.8 \\
Time of adsorption step / s & 989.6 & 1001.0 & 635.6 \\
Time of light reflux step / s & 59.7 & 160.1 & 215.2 \\
Time of heavy reflux step / s & 526.8 & 521.0 & 558.8 \\ \hline
\textbf{Performances} & & & \\ \hline
$CO_2$ purity / \% & 80.0 & 73.1 & 51.8 \\
$CO_2$ recovery / \% & 73.6 & 89.9 & 99.4 \\
Energy consumption / kWh & 125.5 & 210.7 & 320.4 \\
$CO_2$ productivity / (mol/kg.h) & 0.338 & 0.358 & 0.251 \\ \hline
\end{tabular}
\caption{Operation parameters and their corresponding performance for each representative point for the regions of the operation map.}
\label{tab:Mapa}
\end{table}

The duration of each stage for each selected point is highlighted in Figure \ref{Dinamica}. This figure represents the concentration of the column output at the end of a cycle in the CSS.
In Figure \ref{Dinamica}, it is noteworthy that region 1 exhibits significantly higher $CO_2$ concentrations than regions 2 and 3. Specifically, in the depressurization and light reflux stages, the $CO_2$ molar fraction is notably higher when compared to the points in regions 2 and 3, leading to higher $CO_2$ purity. In Figure \ref{Dinamica}, we can also observe that the adsorption step exhibits a higher fraction of $CO_2$ in the column outlet. This indicates that the $N_2$-enriched stream is being contaminated, compromising $CO_2$ purity. Note that at the Pareto front, purity does not reach values higher than 80\%.

\begin{figure}
\centering
\includegraphics[width=14cm]{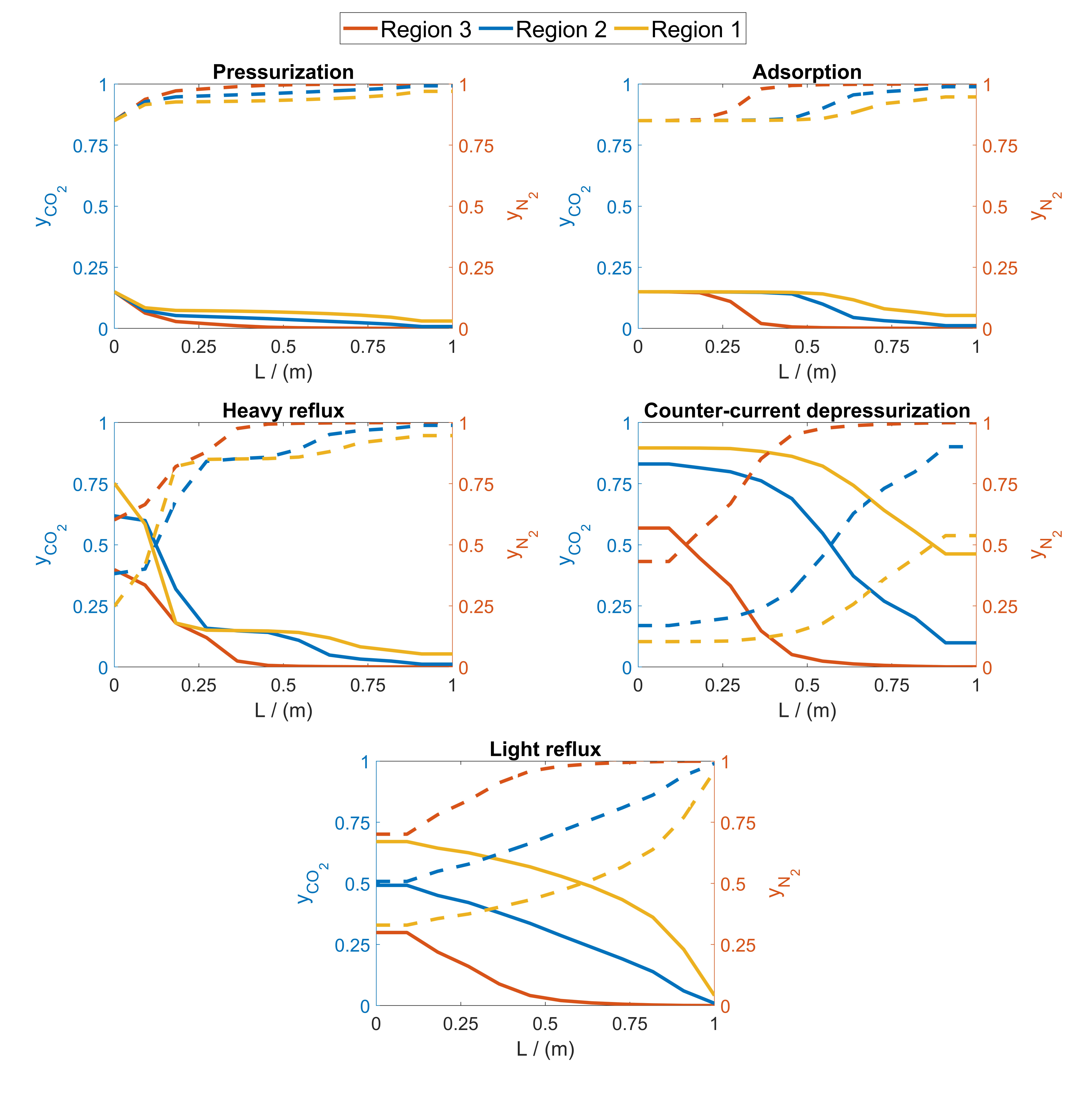}
\caption{Internal bed profiles for  mole fraction of $CO_2$ and $N_2$ (solid line $CO_2$ fraction, dashed line $N_2$ fraction).\label{Perfil}}
\end{figure}

Figure \ref{Perfil} displays the molar fraction profiles of $CO_2$ and $N_2$ along the bed at the end of each stage in the CSS. The aim is to compare these profiles at three distinct points representing each region in the operational map. In region 1, contamination in the enriched $N_2$ flow during adsorption becomes evident. Notably, the $CO_2$ fraction does not reach zero at the upper end of the bed, indicating the persistence of residual $CO_2$ in the enriched $N_2$ stream at the stage exit. This contamination diminishes the efficiency of $CO_2$ recovery. Another important factor is related to region 3, where the recovery of $CO_2$ is prioritized. It is observed that the capture of $CO_2$ occurs only in 50\% of the bed, while the remaining part of the bed adsorbs the lighter component, indicating an underutilization of the column. In region 2, 90\% of the bed is dedicated to capturing $CO_2$, while 10\% captures $N_2$. 

From the standpoint of $CO_2$ purity, at the point corresponding to region 1, there was a more excellent capture of $CO_2$ compared to the other points. This results in a purer $CO_2$ stream at the exit of depressurization, favoring the overall cycle's purity.

The depressurization and light reflux stages are crucial for desorbing the heavy products and regenerating the column adsorption capacity. This is achieved by reducing the pressure in the system, thus, taking advantage of the adsorption isotherm.  The energy consumed in the desorption-related stage primarily depends on the necessity for a vacuum and for how long it is rewuired. 

Thus, increasing recovery in a PSA system can lead to a higher energy demand during the depressurization and purge stages. This extension in the adsorption stage typically implies higher operating times in a high-pressure condition to ensure more efficient retention. 

\begin{figure}
\centering
\includegraphics[width=11 cm]{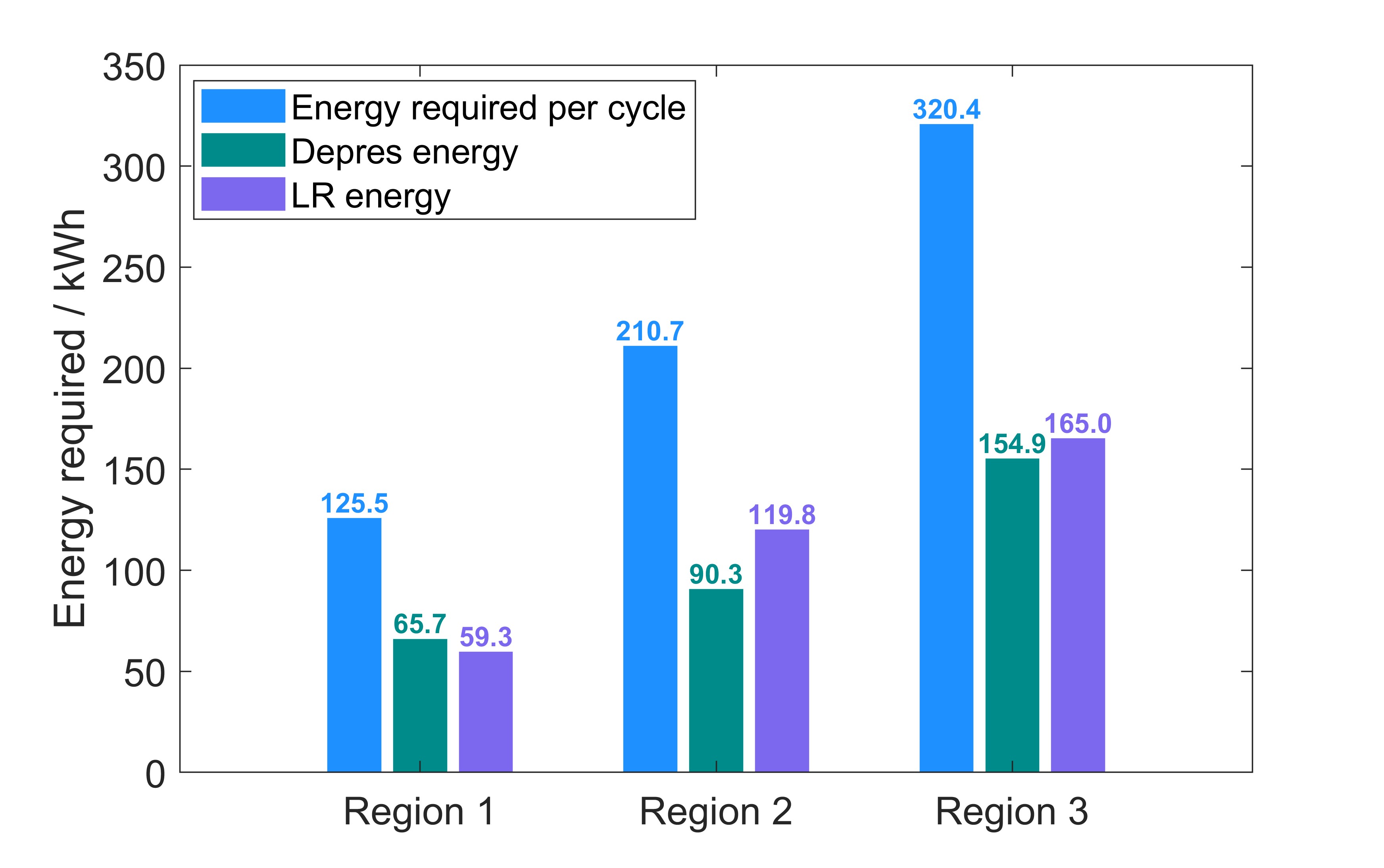}
\caption{Energy required per cycle in the CSS and for the depressurization and light reflux stages for the points representing regions 1, 2, and 3.}
\label{Energia}
\end{figure}

Figure \ref{Dinamica} shows that the points representing regions 2 and 3 depict longer depressurization and light reflux stages than those corresponding to region 1. This pattern also holds for the pressurization and adsorption stages. When the durations of these stages are combined, they yield a higher total time for regions 2 and 3. As a result, there has been an increase in energy requirements, as illustrated in Figure \ref{Energia}.

In Figure \ref{Energia}, we highlight the energies required for the depressurization and light reflux stages, which stand out as the most energy-intensive aspects. The energies required for the other stages are comparatively less significant. 

Another essential discussion concerns the inherent trade-off between purity, recovery, and productivity. To achieve a high purity of $CO_2$, the PSA system is operated to retain $CO_2$ and eliminate $N_2$ selectively. However, this pursuit of purity often results in lower productivity, as a significant portion of the gas feed is discarded to ensure high purity.
On the other hand, when productivity is the focus, the PSA system is optimized to capture as much $CO_2$ as possible without necessarily achieving very high purity. The other side of the trade-off is the fact that recovery refers to the percentage of $CO_2$ captured relative to the total $CO_2$ available in the source. At the same time, productivity concerns the total amount of $CO_2$ captured per cycle time in the CSS (Equation \ref{productivity}).
In general, achieving the desired recovery requires the system to be more efficient in $CO_2$ capture. This often involves the application of stricter operating conditions that ensure a higher purity of $CO_2$ in the final product. However, an increase in recovery often comes at the cost of reduced productivity due to the extended time required for each complete cycle in PSA.
Conversely, when productivity is the primary priority, the system is optimized to capture as much $CO_2$ as possible, even if this means lower recovery.

\begin{figure}
\centering
\includegraphics[width=11 cm]{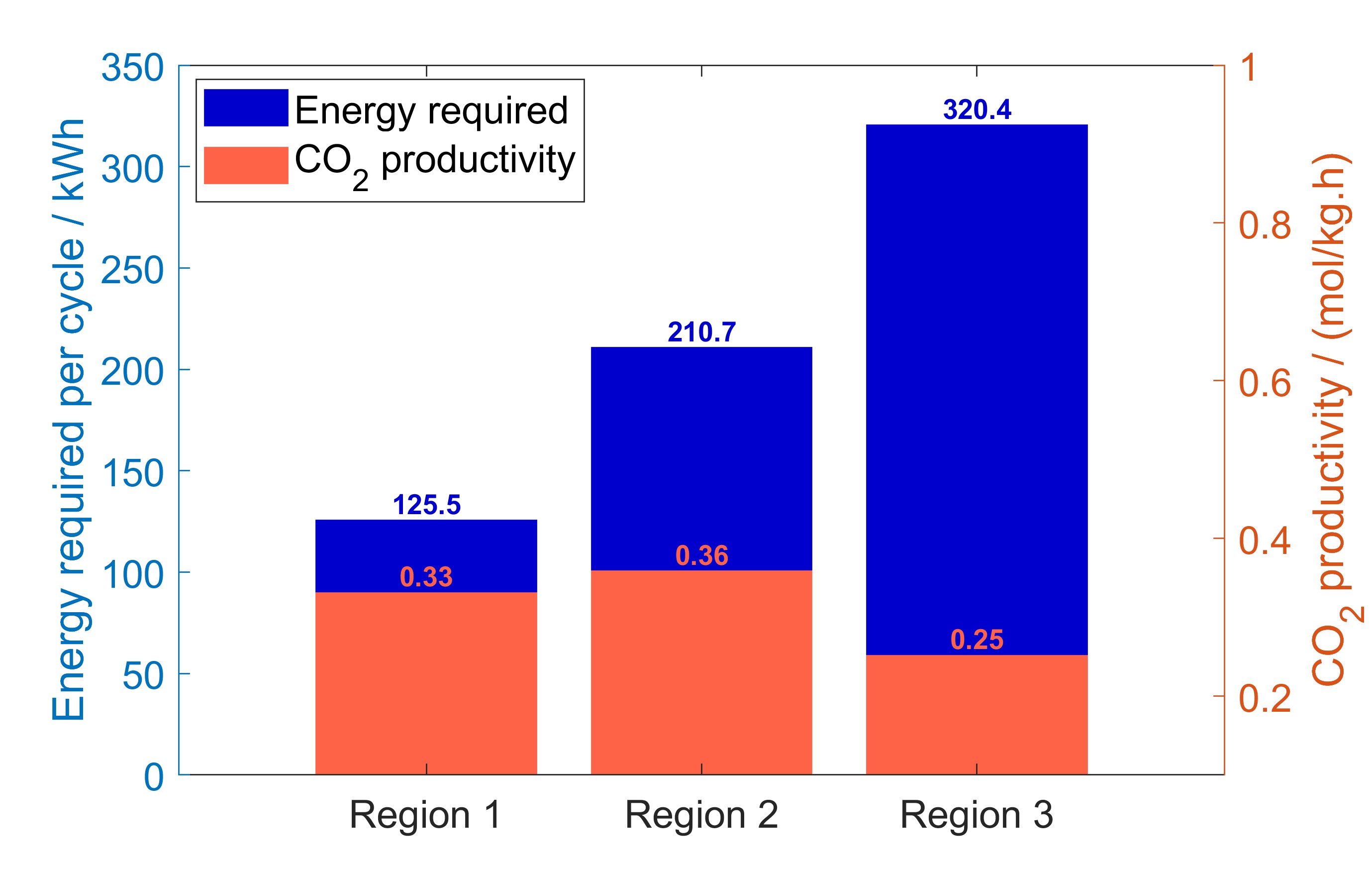}
\caption{Comparison between the energy required and $CO_2$ productivity in the CSS for each representative point in the operating region.}
\label{Energia_prod}
\end{figure}

In Figure \ref{Energia_prod}, productivity is depicted for each point corresponding to regions 1, 2, and 3. Observing this graph makes it striking how it illustrates the trade-off between purity, recovery, and productivity. Regions that require higher purity and recovery often display lower productivity. In contrast, region 2 stands out for achieving a balance between purity and recovery, resulting in a more even productivity. This graph underscores the importance of finding the ideal combination of these factors, considering the specific process goals, to achieve the desired performance in the $CO_2$ capture system.

In Figure  \ref{Energia_prod}, a comparison is also made between the energy required for the three regions and productivity. It is noted that increasing productivity often involves the pursuit of more efficient operating conditions, such as shorter cycle times. However, these improvements often demand more energy, as longer gas compression, adsorbent regeneration, and other process steps may require additional energy. Therefore, an increase in productivity can lead to an increase in energy consumption.

On the other hand, reducing energy consumption is an important goal in many industries, as it lowers operational costs and environmental impacts. However, energy reduction is often associated with decreased productivity since operating processes under milder or less efficient conditions may produce fewer desired components within a given time frame.

Thus, higher $CO_2$ purity (region 1) will require less energy consumption and yield higher productivity, whereas greater recovery (region 3) entails higher energy requirements and lower productivity.

\section{Conclusions}
This study introduces a methodology of surrogated optimization of cyclic adsorption units while offering a robust strategy for assessing the effectiveness of the optimization process. This methodology was evaluated in practice to optimize a pressure swing adsorption unit for $CO_2$ capture.

To do so, we identified four DNN models within a multiple-input, single-output framework. Each of these models was capable of predicting critical process performance parameters. Subsequently, these models were integrated into an optimization framework, utilizing a particle swarm optimizer and a statistical evaluation of the resulting population. The outcome was a comprehensive representation of the Pareto front, including its regions of feasible operation.

To validate the effectiveness of the optimization process, decision variables corresponding to the Pareto front were tested in the phenomenological model, confirming the reliability of the surrogate model. The overlap ensures successful optimization through the substitute model based on the identified DNN, highlighting a significant computational advantage. While the phenomenological model requires substantial time for certain scenarios, the DNN achieves convergence much more efficiently, underscoring the effectiveness of the surrogate-based approach.

Finally, we analyzed the feasible operational domains of the decision variables, accompanied by a correlation map detailing their interplay. This conclusive analysis shed light on the most influential variables shaping process behavior and provided insights into how they can be manipulated to attain specific optimal criteria.

The outcomes of this study offer a significant contribution to decision-making processes. It presents a comprehensive operational map that guides operators in determining the optimal process location and prioritizing specific objectives. This strategic tool enhances the ability to make informed choices, effectively aligning operations with targeted goals.

\section*{Declaration of Competing Interest}

The authors declare that they have no known competing financial interests or personal relationships that could have appeared to influence the work reported in this paper.

\section*{Acknowledgement}
The present work contributes to the completion of a sub-project at SUBPRO, a research-based innovation center within Subsea Production and Processing at the Norwegian University of Science and Technology. The authors would like to express their gratitude for the financial support received from SUBPRO, funded by the Research Council of Norway through grant number 237893, major industry partners, and NTNU.

\FloatBarrier

\bibliographystyle{cas-model2-names}

\bibliography{cas-refs.bib}


\end{document}